\documentclass[]{interact}
\usepackage[utf8]{inputenc}

\usepackage{amssymb}
\usepackage{amsmath}
\usepackage[ruled,vlined]{algorithm2e}
\usepackage{mathtools}
 \usepackage{amsthm}
\usepackage{array}
\usepackage{multirow}
\usepackage{graphicx}
\usepackage{caption}
\usepackage{subcaption}
\usepackage{url}
\usepackage{xcolor}
\usepackage{rotating}
\usepackage[flushleft]{threeparttable}

\theoremstyle{plain}

\theoremstyle{definition}

\usepackage{natbib} 
\setcitestyle{authoryear,open={(},close={)}}
\begin{document}


\title{An ADMM approach for multi-response regression with overlapping groups and interaction effects}
\author{
\name{Theophilus Quachie Asenso\textsuperscript{a*}\thanks{CONTACT Theophilus Quachie Asenso. Email: t.q.asenso@medisin.uio.no}, Manuela Zucknick\textsuperscript{a}  }
\affil{\textsuperscript{a} Oslo Center for Epidemiology and Biostatistics, Institute of Basic Medical Sciences, University of Oslo}
}

\maketitle

\begin{abstract}
In this paper, we consider the regularized multi-response regression problem where there exists some structural relation within the responses and also between the covariates and a set of modifying variables. To handle this problem, we propose MADMMplasso, a novel regularized regression method. This method is able to find covariates and their corresponding interactions, with some joint association with multiple related responses. We allow the interaction term between covariate and modifying variable to be included in a (weak) asymmetrical hierarchical manner by first considering whether the corresponding covariate main term is in the model. For parameter estimation, we develop an ADMM algorithm that allows us to implement the overlapping groups in a simple way. The results from the simulations and analysis of a pharmacogenomic screen data set show that the proposed method has an advantage in handling correlated responses and interaction effects, both with respect to prediction and variable selection performance.

\end{abstract}




\section{Introduction}\label{section1}
The pliable lasso (plasso) \citep{tib} allows one to solve problems involving main effects and their corresponding interaction effects. Let $\mathbf{y}\in \mathbb{R}^{N}$ represent the response from a linear regression model with $N$ observations, $X\in \mathbb{R}^{N\times p}$ be the matrix  of covariates with entries $x_{ij}$ and $Z\in \mathbb{R}^{N\times K}$ be the matrix for the modifying variables with entries $z_{ik}$ for $i=1,\ldots, N$, $j=1,\ldots,p$, $k=1,\ldots,K$. Let $X_j$ be the $j^{th}$ column of $X$, $Z_k$ be the $k^{th}$ column of $Z$ and $\mathbf{1}$ be an $N$-vector of ones. The  pliable lasso model is given as 
\begin{equation}\label{pl1}
 \begin{split}
 \hat{\mathbf{y}}&=\beta_{0}\mathbf{1} +Z\theta_0+\sum_{j=1}^p X_j(\beta_j\mathbf{1}+Z\bm{\theta}_j)\\
 &=\beta_0\mathbf{1} +Z\theta_0+X{\beta}+\sum_{j=1}^p(X_j \circ Z)\bm{\theta}_j,
\end{split} 
 \end{equation}
where ${\beta}\in \mathbb{R}^{p}$, $\theta$ is $p\times K$ matrix of parameters with $j^{th}$ row $\bm{\theta}_j$ and individual entries $\theta_{jk}$, $(X_{j}\circ Z)\in \mathbb{R}^{N\times K}$ is a matrix formed by multiplying each column of $Z$ component-wise
with the column vector $X_j$ and $\theta_0\in \mathbb{R}^{K\times 1}$. The objective function of model \eqref{pl1} is given as 
\begin{multline}\label{pl2}
 M(\beta_0,\theta_0,\beta,\theta)=\frac{1}{2N}\sum_i(y_i-\hat{y}_i)^2 \\
 +  (1-\alpha)\lambda \sum_{j=1}^p (\lVert (\beta_{j},\bm{\theta}_{j})\rVert_2+\lVert \bm{\theta}_{j}\rVert_2)+\alpha\lambda \sum_{j,k}\lvert \theta_{j,k}\rvert,
 \end{multline}
where  $\lambda$ and $\alpha$ are tuning parameters, both $\geq 0$. The model assumes an asymmetric weak hierarchy constraint which is introduced as an overlapping group in the first term of the penalty. The constraints ensure that the interaction term can be nonzero only if the corresponding main term is nonzero. Even though the idea is still young, it has been applied  in different areas, for example to multinomial logistic regression \citep{mplasso}, Cox's proportional hazards model \citep{cox} and support vector machines \citep{svmplasso}. However, in all the above studies, the block-wise coordinate descent procedure was used in solving the problem which includes overlapping groups. The algorithm involves multiple ``if'' statements and a generalized gradient at the final stage. This implies that extending the model to a multi-response case would require rigorous computations like the case of \citeauthor{li2015multivariate} \citeyear{li2015multivariate}, which might be difficult to handle. 

In this paper, we introduce the alternating direction method of multipliers (ADMM) to handle this problem and extend the results from the single response model to a multi-response problem. We provide a publicly available software package \textbf{MADMMplasso} \citep{Asenso_MADMMplasso_2022} implemented in R. We present a brief review on the ADMM algorithm in what follows. 

\subsection{A gentle introduction to ADMM }
Details of the results in this section can be obtained from the works of \cite{boyd2011distributed}, \cite{ye2011efficient}, \cite{deng2013group}, \cite{han2012note}, \cite{Humayoo_2019} and \cite{family}. Here we give a brief introduction to the ADMM algorithm. Given a separable objective function 
\begin{equation}\label{admm1}
    \underset{\beta}{\min} f(\beta)+h(\beta),
\end{equation}
where $f$ and $h$ are both convex, closed and proper, by introducing an auxiliary variable $\omega$, ADMM allows \eqref{admm1} to be re-written as 
\begin{equation}\label{admm2}
    \underset{\beta, \omega}{\min} f(\beta)+h(\omega) \quad \text{s.t} \quad \beta=\omega.
\end{equation}
The problem in \eqref{admm2} can have a corresponding augmented Lagrangian in the form 
\begin{equation}\label{admm3}
    L(\beta,\omega,\gamma)= f(\beta)+h(\omega) + \gamma(\beta-\omega)+ (\rho/2)\lVert \beta-\omega\rVert_2^2,
\end{equation}
where $\rho$ is the augmented Lagrangian parameter or the multiplier. It has been established that the ADMM algorithm converges to the global optimum, under few simple conditions. The algorithm begins with initial parameters $\epsilon^{pri}>0,\epsilon^{dual}>0,\beta^0,\omega^0,\gamma^0$ and  in each  step $t+1$, with $t=1,2,3,\hdots$ indexing the iterations, the ADMM algorithm updates $\beta$ and $\omega$ in an alternating or sequential manner in the following way until the convergence condition is met. 
\begin{equation}
    \begin{split}
    \beta^{t+1} &= \underset{\beta}{\arg\min}\quad L(\beta,\omega^t,\gamma^t)\\
    \omega^{t+1} &= \underset{\omega}{\arg\min}\quad L(\beta^{t+1},\omega,\gamma^t)\\
     \gamma^{t+1} &= \gamma^t+\rho(\beta^{t+1}-\omega^{t+1}).
\end{split}
\end{equation}
The algorithm meets the convergence condition if $s^{t+1}< \epsilon^{dual}$  and $r^{t+1}< \epsilon^{pri}$, where $s^{t+1}=\rho\lVert \omega^{t+1}-\omega^t\rVert_2$ and $r^{t+1}=\lVert \beta^{t+1}-\omega^{t+1}\rVert_2$, representing the dual and primal residuals respectively. The augmented Lagrangian parameter $\rho$ can either be fixed or allowed to change by using the rule: 
\begin{equation}\label{rho}
\rho^{t+1}=\begin{cases}
2\rho^{t} & \text{if} \quad \rho^t> 10s^t\\
\rho^{t}/2 & \text{if} \quad 10\rho^t< s^t\\
\rho^{t} & \text{otherwise} .
\end{cases}
\end{equation} 
The method has received attention (for example in the works of \cite{COFADMM}, \cite{pmlr-v15-ye11a}, \cite{doi:10.1137/120896219}, \cite{han2012note}, \cite{app9204291} and \cite{qin2012structured}) and has been applied to different optimisation problems  including lasso and group lasso and a possibility to include overlapping groups which was introduced in the work of   \cite{deng2013group}. In this paper, we study the possible application of the ADMM algorithm to handle the overlapping groups in the pliable lasso objective function. 

The rest of the paper is organized as follows. In section \eqref{section2}, we present two ADMM algorithms. The first algorithm is an ADMM version of the original pliable lasso linear model. We show that ADMM can handle the overlapping groups in the objective function and will result in the same solution as using the block-wise coordinate descent method in the original pliable lasso paper. The second algorithm is to solve a multi-response problem that has the pliable lasso penalty, and in addition a penalty to consider the correlation between the responses. We assume that the correlation between the response follows a hierarchical tree structure and the responses form groups at various levels within the hierarchy. This implies that each response could belong to several groups, hence the need to allow for overlapping groups. In section \eqref{section3} and \eqref{section4}, we show both the simulations and a real data application, respectively of the proposed method. Section \eqref{section6} is  for the conclusion.

\section{Method}\label{section2}

\subsection{{Model for single response}}
 Let $ B\in\mathbb{R}^{ p\times (K+1)}$ be a matrix of coefficients, with rows and columns indexed from $1$ to $p$ and $1$ to $(K+1)$ respectively, representing the combination of $X$ and $Z$ with the $j^{th}$ row of $\mathbf{\textit{B}}$ defined as $B_j=[\beta_j,\bm{\theta}_j]\in \mathbb{R}^{K+1}$. We can reconstruct the problem in \eqref{pl1} using array notations. Let $W$ be an $ p\times (1+K) \times N$ array constructed as follows for $i=1,\ldots, N$, $j=1,\ldots,p$, $k=1, \ldots, K+1$

\begin{equation}\label{W}
W_{jki}=\begin{cases}
x_{ij}z_{ik} & \text{for} \quad k\neq 1\\
x_{ij} & \text{for} \quad k=1,
\end{cases}
\end{equation} 
then equation \eqref{pl1} is equivalent to the model solution
\begin{equation}\label{new_model1}
\hat{\bm{y}}=\beta_0\mathbf{1} +Z\theta_0+W*\mathbf{\textit{B}},
\end{equation}
where $B$ is the matrix of the coefficients and $W*B$ denotes the $ N$ vector whose $i^{th}$ element takes the form 
\begin{equation}\label{w_tild}
(W*B)_i=\sum_{j=1}^p\sum_{k=1}^{K+1}W_{jki}B_{jk}.
\end{equation}
In equation \eqref{w_tild} above, $W_{ji}=[x_{ij},x_{ij}z_{i1},x_{ij}z_{i2},\ldots, x_{ij}z_{iK}]\in \mathbb{R}^{1\times (K+1)}$. By using \eqref{new_model1},  the objective function of the pliable lasso model in \eqref{pl2} can be rewritten as 

\begin{equation}\label{p1}
\underset{\mathbf{\textit{B}}\in \mathbb{R}^{ p\times (1+K)}}{\min} \frac{1}{2N}\lVert \bm{y-\hat{y}} \rVert_2^2+   (1-\alpha)\lambda \sum_{j=1}^p (\lVert B_{j}\rVert_2+\lVert B_{j(-1)}\rVert_2)  
+\alpha\lambda \sum_{j=1}^p\lVert B_{j(-1)}\rVert_1 .
\end{equation}
Here, the largest group is of the size $K+1$ (to represent $[\beta_j,\bm{\theta}_j]$) and there are two groups (with elements $[\beta_j,\bm{\theta}_j], [\bm{\theta}_j]$). Also, $B_{j(-1)}$ represents  the $j^{\text{th}}$ row of $B$ without the first element i.e, without the coefficient of the main effect $\beta_j$. In what follows in this section, we abuse notation and write $K=K+1$ unless otherwise stated. Let $\{\ B_j^{g_s}\in \mathbb{R}^{k_s} ; s=1,2  \}$ be the groupings of $B_j$, where $g_s$ with $\underset{s}{\cup} g_s \subseteq \{1,2,\ldots, K  \}$ is an index set corresponding to the $s^{th}$ group, and $B_j^{g_s}$, which is a row vector, denotes the subvector of $B_j$ indexed by $g_s$ and $k_s$ represents the size of $g_s$. This implies that $B_j^{g_1}=B_{j}=[\beta_j,  \bm{\theta}_j] \quad \text{and}\quad B_j^{g_2}=B_{j(-1)}=[\bm{\theta}_j]$.
To solve the problem \eqref{p1}, we introduce 2 auxiliary variables as follows; we introduce a matrix  $V\in \mathbb{R}^{p\times \tilde{K}}$, with each $j^{th}$ row vector $V_{j}\in \mathbb{R}^{\tilde{K}}$, defined  for the two groups $(B_j^{g_1},B_j^{g_2})$, and let $\tilde{B}_j=[B_j^{g_1},B_j^{g_2}]\in \mathbb{R}^{\tilde{K}}$, as a row vector. This is constructed  for the overlapping group part of the pliable lasso penalty and let us note that $\tilde{K}=\sum_{s=1}^2 k_s \geq K$ and $V_j^1=B_j^{g_1}$ and $V_j^2=B_j^{g_2}$. The second auxiliary variable, $Q\in\mathbb{R}^{ p\times K} $ is introduced to handle the $l_1$-penalty term.  By introducing these auxiliary variables in problem \eqref{p1}, we  obtain the following optimization problem:

\begin{multline}\label{p2}
\underset{\underset{ V,Q}{ B \in \mathbb{R}^{ p\times K} }, }{\min}\quad \frac{1}{2N}\lVert \bm{y-\hat{y}} \rVert_2^2+   (1-\alpha)\lambda \sum_{j=1}^p\sum_{s=1}^2 \lVert V_{j}^s\rVert_2 +\alpha\lambda \sum_{j=1}^p\lVert Q_j \rVert_1\\
\text{s.t} \quad B_j^{g_s}=V_{j}^s, \quad B_j=Q_j.
\end{multline}
The augmented Lagrangian can be written as 

\begin{multline}\label{lag}
L(B,V, Q,O,P )=\lVert \bm y-\hat{\bm y} \rVert_2^2+   (1-\alpha)\lambda \sum_{j=1}^p\sum_{s=1}^2 \lVert V_{j}^s\rVert_2 +\alpha\lambda \sum_{j=1}^p\lVert Q_j \rVert_1\\
+\sum_j O_j(\tilde{B}_j- V_{j})^T+\langle P,B-Q\rangle + \frac{\rho}{2} \sum_j\sum_s\lVert B_j^{g_s}-V_j^s   \rVert_2^2 +\frac{\rho}{2} \lVert B-Q   \rVert_2^2,
\end{multline}
where $O\in \mathbb{R}^{p\times \tilde{K}} $ and $P\in\mathbb{R}^{p\times K}$ are the Lagrange multipliers associated with $V$ and $Q$ respectively and $\langle A,B\rangle \equiv Tr(A^TB)$ denotes the matrix inner product. 

\subsubsection{Updates for the ADMM algorithm }
In each iteration, the algorithm will update $B, V, Q, P$ and $O$ alternatively until the stopping conditions  are met. We show the various updates in what follows:
\begin{enumerate}
    \item $B_j$ update: We begin by showing how $\tilde{B}_j=[B_j^{g_1},B_j^{g_2}]\in \mathbb{R}^{\tilde{K}}$ is constructed.  Let $\tilde{B}_j=(GB_j^T)^T$ with each row of $G\in \mathbb{R}^{\tilde{K}\times K}$, defined as $G_{(s,l)}^k=1$ for $k\in g_s$ and $0$ elsewhere, with rows indexed by $(s,l)$ such that $s\in \{1,2\} $ and $l\in g_s$. For example given $B\in \mathbb{R}^{5\times 4}$ and knowing that in each $j$ are two groups (with elements $[\beta_j,\bm\theta_j], [\bm\theta_j]$), $G$ will be represented as a $(2\times 4)\times 4$ matrix with elements, 
    \begin{equation}\label{G_matrix}
        \begin{bmatrix} 
	1 & 0 & 0 & 0\\
	0 & 1 & 0& 0\\
	0 & 0 & 1 & 0\\
         0 & 0 & 0 & 1\\
	0 & 0 & 0& 0\\
	0 & 1 & 0 & 0\\
        0 & 0 & 1& 0\\
	0 & 0 & 0 & 1\\
	\end{bmatrix}
  \end{equation}
 From \eqref{G_matrix} above, $\tilde{B}_j$ will have elements $[B_{j1},B_{j2},B_{j3},B_{j4},0,B_{j2},B_{j3},B_{j4}]$, which is the same as the original elements $[\beta_{j},\theta_{j1},\theta_{j2},\theta_{j3},0,\theta_{j1},\theta_{j2},\theta_{j3}]$, which is an expansion of $[\beta_j,\bm\theta_j], [\bm\theta_j]$. The update for  $B_j$ is given as 

\begin{multline}\label{beta_j}
\underset{B_j }{\arg\min}  \frac{1}{2N}\sum_{i=1}^N( y_i-\beta_0-Z_i\theta_0-W_{ji}\mathbf{\mathit{{B}}}_{j}^T-\sum_{t\neq j} W_{ti}\mathbf{\mathit{{B}}}_ t^T  )^2\\   
+ O_j(\tilde{B}_j)^T+ P_j(B_{j})^T + \frac{\rho}{2} \lVert \tilde{B}_j-V_j   \rVert_2^2 +\frac{\rho}{2} \lVert B_{j}-Q_j   \rVert_2^2, \quad t\neq j,
\end{multline}
where $Z_i$ is the $i^{th}$ row of $Z$ with individual elements $Z_{ik}$. Let $\tilde{\bm{y}}=\bm{y}-\beta_0\mathbf{1}_N -Z\theta_0$ and define $\bm{r}_{(-j)}=\tilde{\bm{y}}-\sum_{t\neq j} W_t\mathbf{\mathit{{B}}}_ t$ as the partial residual with the $j^{th}$  group removed, where we have represented $W_j$ as $W_j=[W_{j1}^T:W_{j2}^T:\ldots : W_{jN}^T]^T\in \mathbb{R}^{N\times K}$. Further computations will show that, 

\begin{equation}\label{beta_j_update}
B_ j=\left(\frac{W_j^TW_j}{N}+\rho(G^T G+I)\right)^{-1} \left(\frac{1}{N} W_j^T \bm{r}_{(-j)}+\rho(G^TV_j^T+Q_j^T-G^TO_j^T-P_j^T ) \right) 
\end{equation}

\item $V_j$ update: From \eqref{lag}, we can update $V$ by solving the problem

\begin{equation}\label{V_j1}
\underset{V_j }{\arg\min} \quad (1-\alpha)\lambda \sum_{s=1}^2 \lVert V_{j}^s\rVert_2 
+ O_j(- V_{j})^T + \frac{\rho}{2} \sum_s\lVert B_j^{g_s}-V_j^s   \rVert_2^2.
\end{equation}
This can be simplified  as 
\begin{equation}\label{V_j}
\underset{V_j }{\arg\min} \sum_{s=1}^2\left[ \frac{\rho}{2}\Bigg\|B_j^{g_s}-V_{j}^s+\frac{O_{j}^s}{\rho} \Bigg\|_2^2 +(1-\alpha)\lambda\lVert V_{j}^s\rVert_2 \right].
\end{equation}
Further computations similar to the group lasso problem \citep{li2015multivariate, deng2013group} will show that \eqref{V_j} has a closed form solution by the soft threshold formula 
\begin{equation}
\begin{split}
V_{j}^s&=\max \Bigg\{ \lVert r_s\rVert_2-\frac{(1-\alpha)\lambda}{\rho},0 \Bigg\}\frac{r_s}{\lVert r_s\rVert_2}, \\ r_s&=B_j^{g_s}+O_{j}^s
\end{split}
\end{equation}

\item $Q_j$ update: To update $Q_j$, we should note that we added the main effect $\beta_j$  to the $B_j$ vector in the previous updates. However, since we wish  to avoid the effect of the  $l_1$-norm on it,  we remove the penalty effect here. The $Q$ component of the optimization problem \eqref{lag} is updated as 
\begin{equation}\label{Q_j1}
\underset{Q_j }{\arg\min} \quad  \alpha\lambda \lVert Q_j \rVert_1
+ P_j(-Q_j)^T  +\frac{\rho}{2} \lVert B_j-Q_j   \rVert_2^2,
\end{equation}
which can further be expressed as 
\begin{equation}\label{Q_j}
\underset{Q_j }{\arg\min} \quad  \frac{\rho}{2}\Bigg\|B_j-Q_{j}+\frac{P_{j}}{\rho} \Bigg\|_2^2 +\alpha\lambda\lVert Q_{j}\rVert_1.
\end{equation}
Further computations similar to the lasso problem \citep{lasso,sparse} will show that 
\begin{equation}
\begin{split}
Q_{jk}&=sign\left(B_{jk}+P_{jk}  \right)\left( \Bigg |B_{jk}+P_{jk} \Bigg| -\frac{\alpha\lambda}{\rho} \right)_+,  \quad \text{for} \quad k\neq 1  \\
Q_{j1}&=B_{j1}+P_{j1},   \quad \text{for} \quad k= 1,
\end{split}
\end{equation}
where $(a)_+=\max (a,0)$.

\item {$P$ and $O$ updates}:
The two Lagrange multipliers ($O,P$), associated with the penalties in \eqref{lag} are updated as follows; 

\begin{equation}
\begin{split}
P&\leftarrow P+B-Q\\
O&\leftarrow O+ (GB^T)^T-V
\end{split}
\end{equation}

\end{enumerate}


\subsection{{Extension to multi-response model}}
With the motivation from the model for a single-response variable, we proceed to propose an ADMM method for a multi-response model, where the response variables are assumed to have some correlation structures.  The section begins with the model formulation in what follows. Let $Y\in \mathbb{R}^{N\times D}$ represent the response from a regression model with $N$ number of observations, with each observation having  $D$ responses, $X\in \mathbb{R}^{N\times p}$ be the matrix  of covariates with entries $x_{ij}$ and $Z\in \mathbb{R}^{N\times K}$ be the matrix for the modifying variables with entries $z_{ik}$ for $i=1,\ldots, N$, $j=1,\ldots,p$, $k=1,\ldots,K$. Let $X_j$ be the $j^{th}$ column of $X$, $Z_k$ be the $k^{th}$ column of $Z$ and $\mathbf{1}$ be an $N$-vector of ones. The pliable lasso problem in \eqref{pl1} can be extended to a   multi-response model with interaction effect, where each response $Y_d$ can be modeled as  
\begin{equation}\label{mm1}
\hat{Y}_{d}=\beta_{0d}\mathbf{1}+Z\theta_{0d}+\underset{j}{\sum} X_{j}\beta_{jd}\mathbf{1}+\underset{j}{\sum}(X_{j}\circ Z)\bm\theta_{jd},
\end{equation}
with   $d=1,\ldots,D$ and $(X_{j}\circ Z)\in \mathbb{R}^{N\times K}$ is a matrix formed by multiplying each column of $Z$ component-wise
with the column vector $X_j$. The parameter $\beta_0=(\beta_{01},\beta_{02},\ldots,\beta_{0D})^T$ is a $D$ vector of intercepts for the main effect,  $\theta_0\in \mathbb{R}^{K\times D}$ is the matrix of coefficients for the modifying variable $Z$ with each column represented by $\theta_{0d}\in \mathbb{R}^K$, $\beta$ is a $p\times D$ matrix of coefficients for the main effect with each row represented by $\beta_j$, having individual entries $\beta_{jd}$, $\theta$ is a $p\times K \times D$ array of size $D$ with each $d=1,\ldots , D$ element having $p\times K$ matrix where each $j^{th}$ row of the $d^{th}$ matrix is represented by $\bm\theta_{jd}$.   Let $B\in \mathbb{R}^{p\times (K+1)\times D}$ be an array of size $D$ with each $d=1,\ldots, D$ representing a matrix of coefficients for the main and interaction effects for response $d$, with rows and columns indexed from $1$ to $p$ and $1$ to $(K+1)$, respectively.  This is to cater for the combination of $X$ and $Z$ with the $j^{th}$ row of $B_d$ defined as $\mathbf{\textit{B}}_{jd}=[\beta_{jd},\bm\theta_{jd}]\in \mathbb{R}^{K+1}$. We can reconstruct the equation \eqref{mm1} using array notations. Let $W$ be an $ p\times (1+K) \times N$ array with same construction as \eqref{W}, then the model for all responses in equation \eqref{mm1} is equivalent to the model 
\begin{equation}
\hat{Y}=\mathbf{1}\beta_0^T +Z\theta_0+[W*B_1: W*B_2:\ldots:W*B_D].
\end{equation}
We will abuse notation to define   $W*B=[W*B_1: W*B_2:\ldots:W*B_D]$ to denote the $N\times D$ matrix whose $id$ element takes the form 
\begin{equation}\label{w_tild1}
(W*B)_{id}=\sum_{j=1}^p\sum_{k=1}^{K+1}W_{jki}B_{jkd}, \quad i=1,\ldots N, \quad d=1,\ldots, D.
\end{equation} 
The objective function for a general multi-response pliable lasso model can be written as 
\begin{multline}\label{pp1}
\underset{B\in \mathbb{R}^{ p\times (1+K)\times D}}{\min} \quad \frac{1}{2N}\lVert Y-\hat{Y} \rVert_F^2 \\+ \sum_{d=1}^D \left[ (1-\alpha)\lambda \sum_{j=1}^p (\lVert B_{jd}\rVert_2+\lVert B_{j(-1)d}\rVert_2)  
+\alpha\lambda \sum_{j=1}^p\lVert B_{j(-1)d}\rVert_1 \right].
\end{multline}
Here also, the largest group is of the size $K+1$ (to represent $[\beta_{jd},\bm\theta_{jd}]$) and there are two groups (with elements $[\beta_{jd},\bm\theta_{jd}], [\bm\theta_{jd}]$). Also, $B_{j(-1)d}$ represents  the $j^{\text{th}}$ row of $\mathbf{\textit{B}}_d$ without the first element i.e, without the coefficient of the main effect $\beta_{jd}$. 

\subsubsection{Multi-response pliable lasso with tree-guided structure}
The problem \eqref{pp1} can be solved as a single-response problem with interaction effects where each of the responses could be treated as independent of the others.  However, in our case, we assume correlation among the response variables. To handle this correlation, we allow the response variables to form  overlapping groups and having a structure, specifically a tree-like structure in this paper. More details about the tree-like structure can be found in the paper ``Tree-guided group lasso for multi-response regression with structured sparsity, with an application to EQTL mapping" \citep{tree} and similar structures can also be found in \cite{li2015multivariate} and in \cite{zhao2020structured}. 

We represent  the set of internal and leaf nodes of the tree as $M_{\text{int}}$, $M_{\text{leaf}}$ of size $|M_{\text{int}}|$ and $|M_{\text{leaf}}|$ respectively, the group of responses forming an internal node $m\in M_{\text{int}}$ as $\mathcal{G}_m$, where $\mathcal{G}_m\subseteq \{1,\ldots, D\}$ and let $B_j^{\mathcal{G}_m}$ denotes the $j^{th}$ sub-vector of $B$, indexed by $\mathcal{G}_m$ with a group weight $w_m$. Each sub-vector $B_j^{\mathcal{G}_m}$ has elements $\{B_{jd}; d\in \mathcal{G}_m \}$. Adding the overlapping groups in the responses to the problem \eqref{pp1} gives us

\begin{multline}\label{pp2}
\underset{B\in \mathbb{R}^{p\times (1+K)\times D}}{\min} \frac{1}{2N}\lVert Y-\hat{Y} \rVert_F^2   +\lambda_1 \sum_{j=1}^p \sum_{m\in M_{\text{int}}} w_m\lVert B_j^{\mathcal{G}_m}\rVert_2 +\lambda_2 \sum_{j=1}^p \sum_{m\in M_{\text{leaf}}} w_m\lVert B_j^{\mathcal{G}_m}\rVert_2 \\ + \sum_d^D \left[ (1-\alpha)\lambda_3 \sum_{j=1}^p (\lVert B_{jd}\rVert_2+\lVert B_{j(-1)d}\rVert_2)  
+\alpha\lambda_3 \sum_{j=1}^p\lVert B_{j(-1)d}\rVert_1 \right].
\end{multline}
%
The first penalty terms  ensure that correlated responses are penalized within groups while ensuring sparsity among the coordinate $j$ and the second penalty term is the pliable lasso part to ensure that main and interaction effects are considered in a hierarchical format.

To solve the problem \eqref{pp2} by the ADMM approach, we introduce four auxiliary variables as follows.  Let $E\in \mathbb{R}^{( \sum_{m\in M} \lvert \mathcal{G}_m\rvert)\times (p+p\times K)}$ be the matrix with each $E_j\in \mathbb{R}^{ \sum_{m\in M} \lvert \mathcal{G}_m\rvert}\times (K+1)$ to represent the  groups formed by internal nodes $M$ on the $j^{th}$ coordinate , and $E^{\mathcal{G}_m}\in \mathbb{R}^{\lvert \mathcal{G}_m\rvert \times (p+p\times K)}$ being the block representation of group $\mathcal{G}_m$ and equal to $\mathit{B}^{\mathcal{G}_m}$. This is defined for the overlapping groups in the responses for $(\mathit{B}_j^{\mathcal{G}_1},\mathit{B}_j^{\mathcal{G}_2},\ldots, \mathit{B}_j^{\mathcal{G}_M})$ with $\mathit{\tilde{\tilde{B}}}_j\equiv [\mathit{B}_j^{\mathcal{G}_1 T},\mathit{B}_j^{\mathcal{G}_2 T},\ldots, \mathit{B}_j^{\mathcal{G}_M T} ]^T\in \mathbb{R}^{\sum_{m\in M} \lvert \mathcal{G}_m\rvert\times (K+1) }$. Note that we have defined $E$ to contain both the main effect and the interaction effect of each $j$, hence each row of $E$ can be seen as having $p(K+1)$ columns. We also introduce  an array $\tilde{E}\in\mathbb{R}^{ p\times (1+K)\times D} $ to handle the $l_2$-penalty of the leaf node ($\lambda_2 \sum_{j=1}^p \sum_{m\in M_{\text{leaf}}} w_m\lVert B_j^{\mathcal{G}_m}\rVert_2$) in \eqref{pp2}.

In what follows in this section, we abuse notation and write $K=K+1$ unless otherwise stated. Let $\{\ B_{jd}^{g_s}\in \mathbb{R}^{k_s} ; s=1,2  \}$ be the grouping of elements of $B_{jd}$, where $g_s$ with $\underset{s}{\cup} g_s \subseteq \{1,\ldots, K  \}$ is an index set corresponding to the $s^{th}$ group, and $B_{jd}^{g_s}$ which is a row vector, denotes the subvector of $B_{jd}$, indexed by $g_s$ and $k_s$ represents the size of $g_s$. We  introduce the second auxiliary variable as an array $V$ of size $D$ with each of the index $d=1,\ldots, D$ having a matrix $V_d \in \mathbb{R}^{p\times \tilde{K}}$, with each $j^{th}$ row vector $V_{jd}\in \mathbb{R}^{\tilde{K}}$, defined  for the two groups $(B_{jd}^{g_1},B_{jd}^{g_2})$.  This is  for the overlapping part of the pliable lasso penalty and let us note that $\tilde{K}=\sum_{s=1}^2 k_s \geq K$ and $V_{jd}^1=B_{jd}^{g_1}$ and $V_{jd}^2=B_{jd}^{g_2}$. Also, let $\tilde{B}_{jd}\equiv [B_{jd}^{g_1},B_{jd}^{g_2}]\in \mathbb{R}^{\tilde{K}}$ and $\tilde{B}_{d} \in \mathbb{R}^{p\times \tilde{K}}$ be formed with elements from $\tilde{B}_{jd}$ such that each row is $\tilde{B}_{jd}$. 

 Finally, we introduce an array $Q\in\mathbb{R}^{D\times (p\times K}) $ to handle the $l_1$-penalty of the pliable lasso.  Implementing these auxiliary variables on problem \eqref{pp2} leads to the following;

\begin{multline}\label{mad1}
\underset{B, E,\tilde{E}, V,Q   }{\min} \quad \frac{1}{2N}\lVert Y-\hat{Y} \rVert_F^2+\lambda_1\sum_{j=1}^{p}\sum_{m\in M_{\text{int}}} w_{m}\lVert E_{j}^{\mathcal{G}_m} \rVert_2 +\lambda_2 \sum_{j=1}^p \sum_{m\in M_{\text{leaf}}} w_m\lVert \tilde{E}_j^{\mathcal{G}_m}\rVert_2\\  +\sum_d(1-\alpha)\lambda_3 \sum_{j=1}^p\sum_{s=1}^2 \lVert V_{jd}^s\rVert_2 +\alpha\lambda_3 \sum_{j=1}^p\lVert Q_{jd} \rVert_1\\
\text{s.t}  \quad B_{j}^{\mathcal{G}_m} =E_{j}^{\mathcal{G}_m}, \quad  B_d =\tilde{E}_d, \quad B_{jd}^{g_s}=V_{jd}^s, \quad B_d=Q_d.
\end{multline}
In the above,  $\tilde{E}_j^{\mathcal{G}_m}$ is the $j^{th}$ component of each $\tilde{E}_d$ and since it is a leaf node, each will have a single element. Therefore the penalty $\lambda_2 \sum_{j=1}^p \sum_{m\in M_{\text{leaf}}} w_m\lVert \tilde{E}_j^{\mathcal{G}_m}\rVert_2$ can be rewritten as $\lambda_2 \sum_d\sum_{j=1}^p  w_d\lVert \tilde{E}_{jd}\rVert_2$ which is just a further shrinkage on each $j$.
The augmented Lagrangian form of \eqref{mad1} can be written as 
\begin{multline}\label{mad2}
{L}(B,E,\tilde{E},V,Q,H,\tilde{H},O,P)= \frac{1}{2N}\lVert Y-\hat{Y} \rVert_F^2+\lambda_1\sum_{j=1}^{p}\sum_{m\in M_{\text{int}}}w_{m}\lVert E_{j}^{\mathcal{G}_m} \rVert_2 \\+\lambda_2 \sum_{d}\sum_{j=1}^p  w_d\lVert \tilde{E}_{jd}\rVert_2    +\sum_d(1-\alpha)\lambda_3 \sum_{j=1}^p\sum_s \lVert V_{jd}^s\rVert_2 +\alpha\lambda_3 \sum_{j=1}^p\lVert Q_{jd} \rVert_1\\
+\sum_j H_j(\tilde{\tilde{B}}_{j}-E_{j})^T+\sum_d\langle \tilde{H}_d,B_d-\tilde{E}_d\rangle +\sum_d\sum_j O_{jd}(\tilde{B}_{jd}-V_{jd})^T+\sum_d\langle P_d,B_d-Q_d\rangle\\
+\frac{\rho}{2}\sum_j\lVert\tilde{\tilde{B}}_{j}-E_{j}\rVert_2^2 +\frac{\rho}{2}\sum_d\lVert B_d-\tilde{E}_d\rVert_2^2+\frac{\rho}{2}\sum_d\sum_j\sum_s\lVert \tilde{B}_{jd}^s-V_{jd}^s \rVert_2^2+\frac{\rho}{2}\sum_d\lVert B_d-Q_d\rVert_2^2.
\end{multline}
In the problem \eqref{mad2} above, $H,\tilde{H},O,P$ have the same dimensions as $E,\tilde{E}, V$ and $Q$ respectively, and  are called the Lagrange multipliers and $\rho$ is a penalty  parameter. 

\subsubsection{Update for the ADMM algorithm}
Just like for the single-response problem,  the algorithm will update $B, E,\tilde{E}, V, Q, H,\tilde{H}, P$ and $O$ alternatively in each iteration until the stopping criteria are met. We show the various updates in what follows:
\begin{enumerate}
    \item $B$ update: We begin by showing the construction of $\tilde{B}_d$ and $\tilde{\tilde{B}}$. Let $\tilde{B}_{d}=(GB_{d}^T)^T$ with each row of $G\in \mathbb{R}^{\tilde{K}\times K}$, defined as $G_{(s,l)}^k=1$ for $k\in g_s$ and $0$ elsewhere, with rows indexed by $(s,l)$ such that $s\in \{1,2\} $ and $l\in g_s$. Also, let $\tilde{\tilde{B}}=IB^T$ with $I\in \mathbb{R}^{( \sum\lvert\mathcal{G}_m\rvert)\times D}$, and $I_{m,u}^d= w_m$ for $d\in \mathcal{G}_m$ and $0$ otherwise, indexed by $(m,u)$ such that $m\in M $ and $u\in \mathcal{G}_m$. An example of these two constructions is shown in \eqref{G_matrix}. Note that we have written the matrix $B\in \mathbb{R}^{(p+p\times K)\times D}$ in the construction of $\tilde{\tilde{B}}$ above as a transformation of the array $B$ such that, each column $B_d$  (of matrix $B$) is a vectorized version of the matrix $B_d$ of the array $B$.   The $B$ component of \eqref{mad2} is updated as follows;
\begin{multline}\label{beta_j_d}
\underset{B  }{\arg\min} \quad \frac{1}{2N}\lVert Y-\mathbf{1}\beta_0^T-Z\theta_0-W*B \rVert_F^2\\+ H(\tilde{\tilde{B}})^T +\sum_d \tilde{H}_d(B_{d})^T +\sum_d O_d(\tilde{B}_{d})^T+\sum_d P_d(B_{d})^T\\
+\frac{\rho}{2}\lVert \tilde{\tilde{B}}-E\rVert_2^2 +\frac{\rho}{2}\sum_d\lVert B_{d}-\tilde{E}_d\rVert_2^2+\frac{\rho}{2}\sum_d\lVert \tilde{B}_{d}-V_d \rVert_2^2+\frac{\rho}{2}\sum_d\lVert B_{d}-Q_d\rVert_2^2.
\end{multline}
Let $\tilde{W}$ denote the $N\times (p+p\times K)$-dimensional matrix of $W$ and 
let $R=Y-\mathbf{1}\beta_0^T-Z\theta_0$, with each column $R_d,$ being an $N$ column vector. Taking the derivative of \eqref{beta_j_d} with respect to $B_d$ and equating to zero gives 
\mbox{}
\begin{equation}\label{beta_j_d_update}
\begin{split}
B_d&=\frac{ \phi }{\frac{\tilde{W}^T\tilde{W}}{N}+C_d },\\
    \text{where}\quad \phi= &\frac{1}{N}\tilde{W}^TR_d+(\rho(I^T E-I^T H))_d \\ & +vec( \rho(EE_d-HH_d ) (\rho(G^T(V_d)^T - G^T(O_d)^T ))^T  +\rho(Q_d-P_d ) ), 
    \end{split}
\end{equation}
and we have allowed  $\frac{a}{b}$ to represent $(b)^{-1}a$ for the sake of simplicity and $vec(a)$ is a vectorized form of $a$. The resulting $B_d$ which is a vector, is then transformed into a matrix of dimension $p\times (K)$.  To construct $C$ in \eqref{beta_j_d_update}, we first defined $\tilde{I}$ as the diagonal of $I^TI$ and $\tilde{G}$ to be the diagonal of $G^TG$ which has elements $\{1,2,2,2,\ldots,2\}$. Then we defined  $C$ to be an array of size $D$ with each $d=1,\ldots, D$ being a $(p+p\times K)\times (p+p\times K)$ diagonal matrix with the first $p$ elements equal to $\rho(1+\tilde{G}_1)+\rho(\tilde{I}_d+1)$ and the $p+1$ to $p+p\times K$ elements equal to $3\rho+\rho(\tilde{I}_d+1)$ since the elements of $\tilde{G}_j=2, j=2,\ldots, K$. This construction is to cater for the groupings within the responses and within  each coordinate $j=1,\ldots, p$. 

\item $V_{jd}$ update: from \eqref{mad2}, we update $V$ by solving the problem
\begin{multline}\label{V_dj1}
\underset{V_{jd} }{\arg\min}  \quad  (1-\alpha)\lambda_3 \sum_s \lVert V_{jd}^s\rVert_2 +  O_{jd}(-V_{jd})^T
+\frac{\rho}{2}\sum_s\lVert \tilde{B}_{jd}^s-V_{jd}^s \rVert_2^2.
\end{multline}
With some few computations, this can be simplified as 
\begin{equation}\label{V_dj}
\underset{V_{jd} }{\arg\min} \sum_{s=1}^2\left[ \frac{\rho}{2}\Bigg\|\tilde{B}_{jd}^s-V_{jd}^s+\frac{O_{jd}^s}{\rho} \Bigg\|_2^2 +(1-\alpha)\lambda_3\lVert V_{jd}^s\rVert_2 \right].
\end{equation}
Further computations similar to the group lasso problem will show that, 
\begin{equation}
\begin{split}
V_{jd}^s&=\max \Bigg\{ \lVert r_s\rVert_2-\frac{(1-\alpha)\lambda_3}{\rho},0 \Bigg\}\frac{r_s}{\lVert r_s\rVert_2}, \\ r_s&=\tilde{B}_{jd}^s+O_{jd}^s
\end{split}
\end{equation}

\item $Q_{jd}$ update: 
The $Q$ component of the optimization problem \eqref{mad2} is updated as 

\begin{equation}\label{Q_dj1}
\underset{Q_{jd} }{\arg\min} \quad   \alpha\lambda_3 \lVert Q_{jd} \rVert_1
 + P_{jd}(-Q_{jd})^T+\frac{\rho}{2}\lVert B_{jd}-Q_{jd}\rVert_2^2,
\end{equation}
which can further be expressed as 
\begin{equation}\label{Q_dj}
\underset{Q_{jd} }{\arg\min} \quad \frac{\rho}{2}\Bigg\|B_{jd}-Q_{jd}+\frac{P_{jd}}{\rho} \Bigg\|_2^2 + \alpha\lambda_3\lVert Q_{jd}\rVert_1.
\end{equation}
Further computations, similar to the lasso problem will show that, 
\begin{equation}
\begin{split}
Q_{jkd}&=sign\left(B_{jkd}+P_{jkd}  \right)\left( \Bigg |B_{jkd}+P_{jkd} \Bigg| -\frac{\alpha\lambda_3}{\rho} \right)_+,  \quad \text{for} \quad k\neq 1  \\
Q_{j1d}&=B_{j1d}+P_{j1d},   \quad \text{for} \quad k= 1.
\end{split}
\end{equation}
\item $E_j^{\mathcal{G}_m}$ update: The $E$ component of the optimization problem \eqref{mad2} is updated as 
\begin{equation}\label{E_j1}
\underset{E_j^{\mathcal{G}_m} }{\arg\min} \quad \lambda_1w_{m}\lVert E_{j}^{\mathcal{G}_m} \rVert_2  
+ H_j^{\mathcal{G}_m}(-E_{j}^{\mathcal{G}_m})^T
+\frac{\rho}{2}\lVert \tilde{\tilde{B}}_{j}^{\mathcal{G}_m}-E_{j}^{\mathcal{G}_m}\rVert_2^2,
\end{equation}
which can be simplified as 
\begin{equation}\label{E_j}
\underset{E_j^{\mathcal{G}_m} }{\arg\min} \left[ \lambda_1w_{m}\lVert E_j^{\mathcal{G}_m}\rVert_2+\frac{\rho}{2}\Bigg\|\tilde{\tilde{B}}_j^{\mathcal{G}_m}-E_j^{\mathcal{G}_m}+\frac{H_j^{\mathcal{G}_m}}{\rho} \Bigg\|_2^2 \right].
\end{equation}
Further computations, by following the group lasso example will show that, 
\begin{equation}
\begin{split}
E_j^{\mathcal{G}_m}&=\max \Bigg\{ \lVert r_m\rVert_2-\frac{\lambda_1w_{m}}{\rho},0 \Bigg\}\frac{r_m}{\lVert r_m\rVert_2}, \\ r_m&=\tilde{\tilde{B}}_j^{\mathcal{G}_m}+H_j^{\mathcal{G}_m}
\end{split}
\end{equation}
\item $\tilde{E}_{jd}$ update: The $\tilde{E}$ component of the optimization problem \eqref{mad2} is updated as 
\begin{equation}\label{EE_j1}
\underset{\tilde{E}_{jd} }{\arg\min} \quad \lambda_2w_{d}\lVert \tilde{E}_{jd} \rVert_2  
+ \tilde{H}_{jd}(-\tilde{E}_{jd})^T
+\frac{\rho}{2}\lVert B_{jd}-\tilde{E}_{jd}\rVert_2^2,
\end{equation}
which can be simplified as 
\begin{equation}\label{EE_j}
\underset{\tilde{E}_{jd} }{\arg\min} \left[ \lambda_2w_{d}\lVert \tilde{E}_{jd}\rVert_2+\frac{\rho}{2}\Bigg\|B_{jd}-\tilde{E}_{jd}+\frac{\tilde{H}_{jd}}{\rho} \Bigg\|_2^2 \right].
\end{equation}
Further computations, by following the group lasso example will show that, 
\begin{equation}
\begin{split}
\tilde{E}_{jd}&=\max \Bigg\{ \lVert r_d\rVert_2-\frac{\lambda_2w_{d}}{\rho},0 \Bigg\}\frac{r_d}{\lVert r_d\rVert_2}, \\ r_d&=B_{jd}+\tilde{H}_{jd}
\end{split}
\end{equation}

\item $H,\tilde{H}, P$ and $O$ updates: The three Lagrange multipliers, associated with the penalties in \eqref{mad2} are updated as follows; 

\begin{equation}
\begin{split}
P_d&= P_d+B_d-Q_d\\
O_d&=O_d+ \tilde{B}_d-V_d\\
H&=H+\tilde{\tilde{B}}-E\\
\tilde{H}_d&=\tilde{H}_d+B_d-\tilde{E}_d
\end{split}
\end{equation}

\end{enumerate}

%

\section{Simulations}\label{section3}
This section entails the simulation studies for our proposed model. In the first part, we show the results for the single-response problem. We compare the ADMM pliable lasso algorithm to the original pliable lasso implementation, which uses the block-wise coordinate descent (BCD) method. The second part of the simulation studies is dedicated to the multi-response problem. We compare our proposed method with the tree lasso and the original pliable lasso which assumes that the multiple responses are independent. For the original pliable lasso, we allowed each response to be modeled one at a time, since there is no multi-response version implemented. 

\subsection{Simulation of a single-response problem}
\begin{figure}[ht]
\begin{subfigure}{.5\textwidth}
  \centering
  \includegraphics[height=4cm,width=\textwidth]{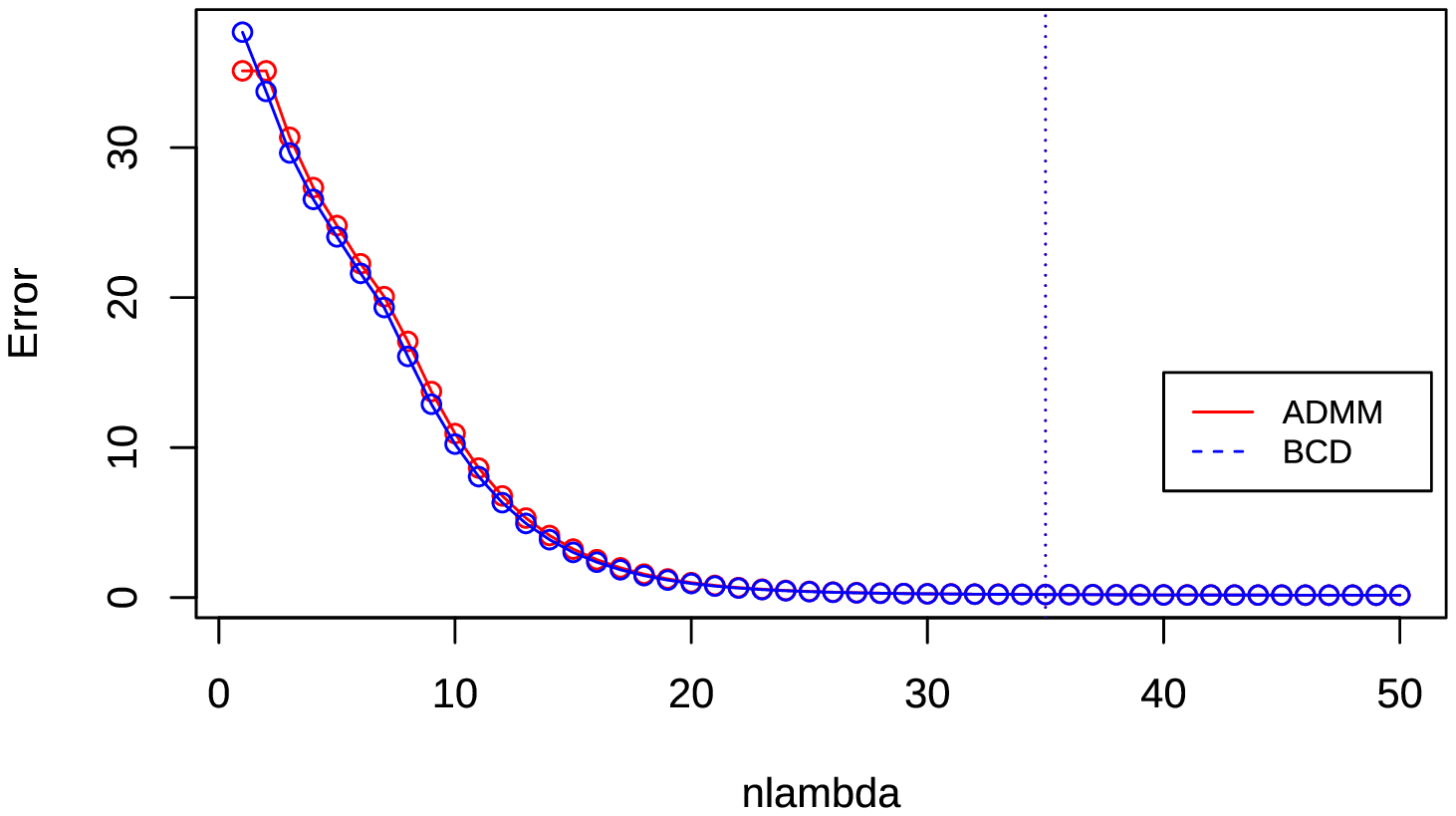}
 \caption{}
\end{subfigure}%
\begin{subfigure}{.5\textwidth}
  \centering
  \includegraphics[height=4cm,width=\textwidth]{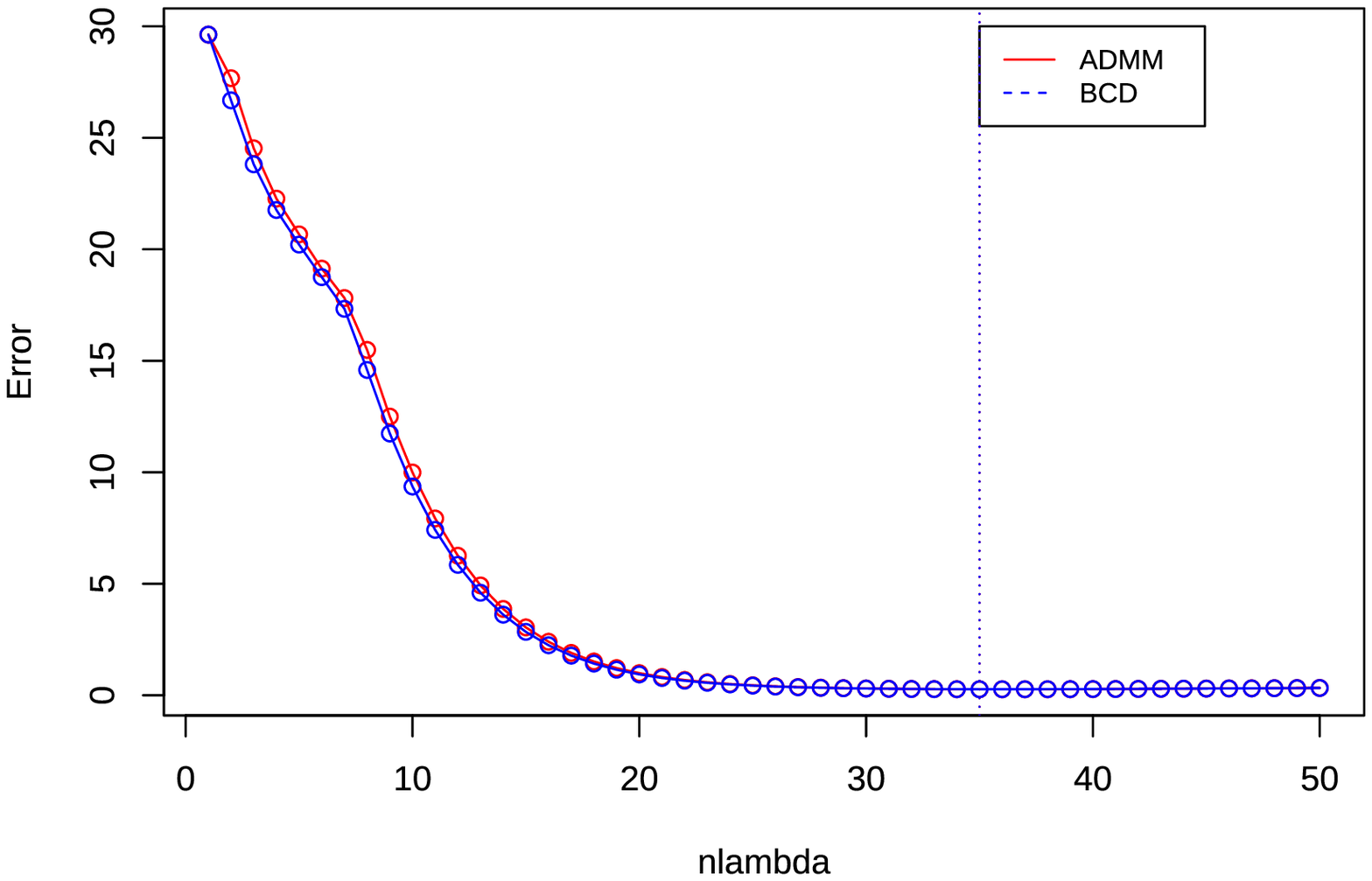}
 \caption{}
\end{subfigure}
\begin{subfigure}{.5\textwidth}
  \centering
  \includegraphics[height=4cm,width=\textwidth]{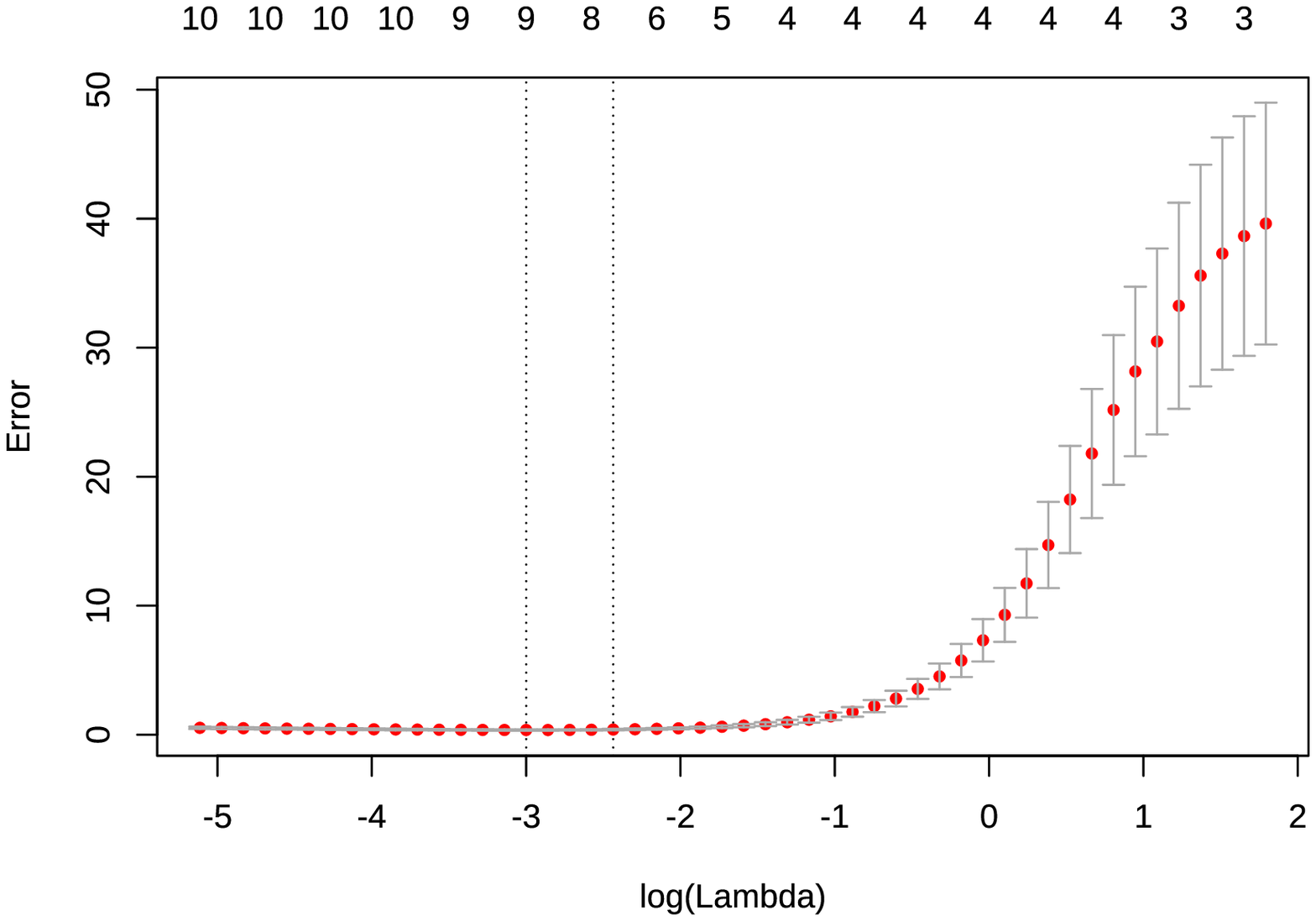}
 \caption{}
\end{subfigure}
\begin{subfigure}{.5\textwidth}
  \centering
  \includegraphics[height=4cm,width=\textwidth]{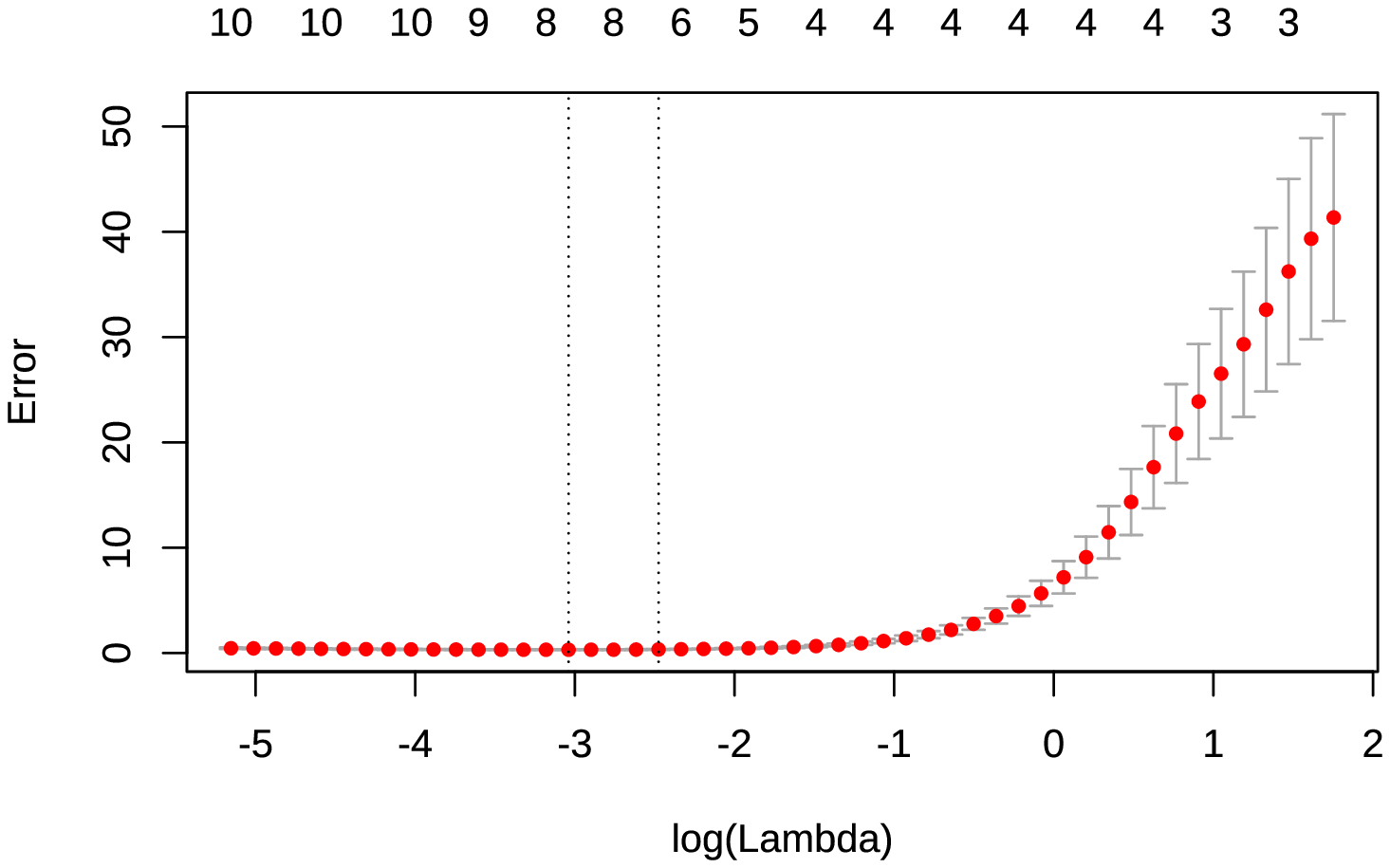}
 \caption{ }
\end{subfigure}
\caption{Plots for training (a), testing (b) and validation error (c-d) for the two methods. Plot (c) and (d) are for the ADMM and BCD respectively.  The error is the mean squared error and ``nlambda" is the length of the $\lambda$ vector. The numbers across the top of the validation error plots (c) and (d) represent the numbers of non-zero main effects at the given $\lambda$ and the vertical broken lines show the selected $\lambda$ from cross-validation. }\label{plot1}
\end{figure}

 We generated data with $N=100$, $p=10$ or $p=50$, $K=3$ and standard Gaussian independent predictors. The generative model is given as  \begin{equation}\label{sim_model1}
y=X_1(\beta_{1}\mathbf{1}+2Z_3)+X_2\beta_{2}+X_3(\beta_{3}\mathbf{1}+2Z_1)+X_4(\beta_{4}\mathbf{1}-2Z_2)+.5\varepsilon.
\end{equation}
Here, $\beta=(2,-2,2,2,0,0,\ldots)$ and $Z$ is $N\times K$ matrix such that $Z_1,\ldots, Z_K \overset{iid}{\sim}\mathbb{N}(0,1)$ and $\varepsilon \sim \mathbb{N}(0,1)$.   The training, validation and test error plots are shown in figure \eqref{plot1}. We find that both the proposed approach and the original plasso, which uses the coordinate descent method, give similar results. We should note here that our main purpose for the single response ADMM approach is to show how the algorithm can be used to effectively handle the overlapping group penalty in the problem.

\begin{figure}[ht]
\begin{subfigure}{.5\textwidth}
  \centering
  \includegraphics[width=\textwidth]{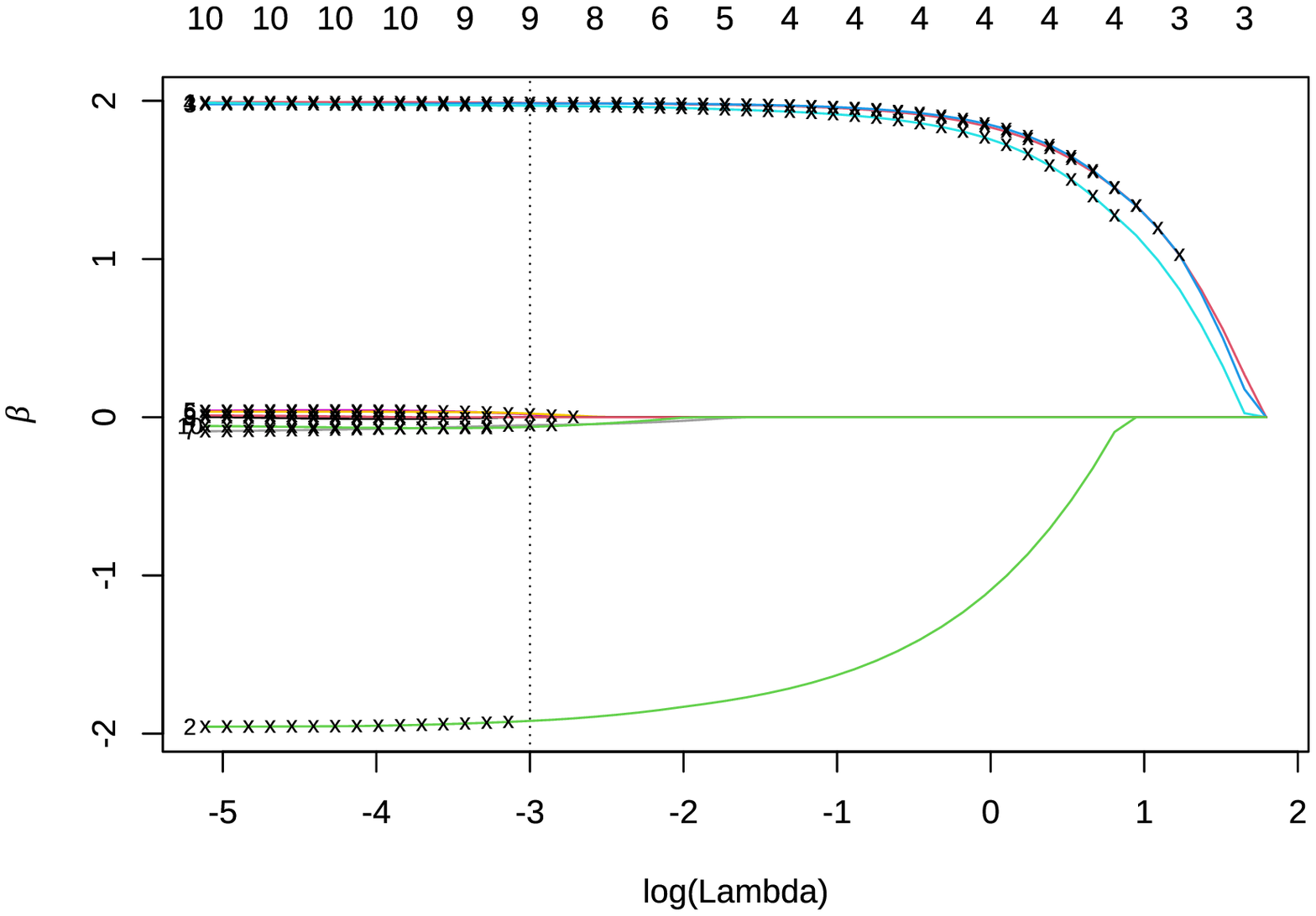}
 \caption{  }\label{adcoeff}
\end{subfigure}%
\begin{subfigure}{.5\textwidth}
  \centering
  \includegraphics[width=\textwidth]{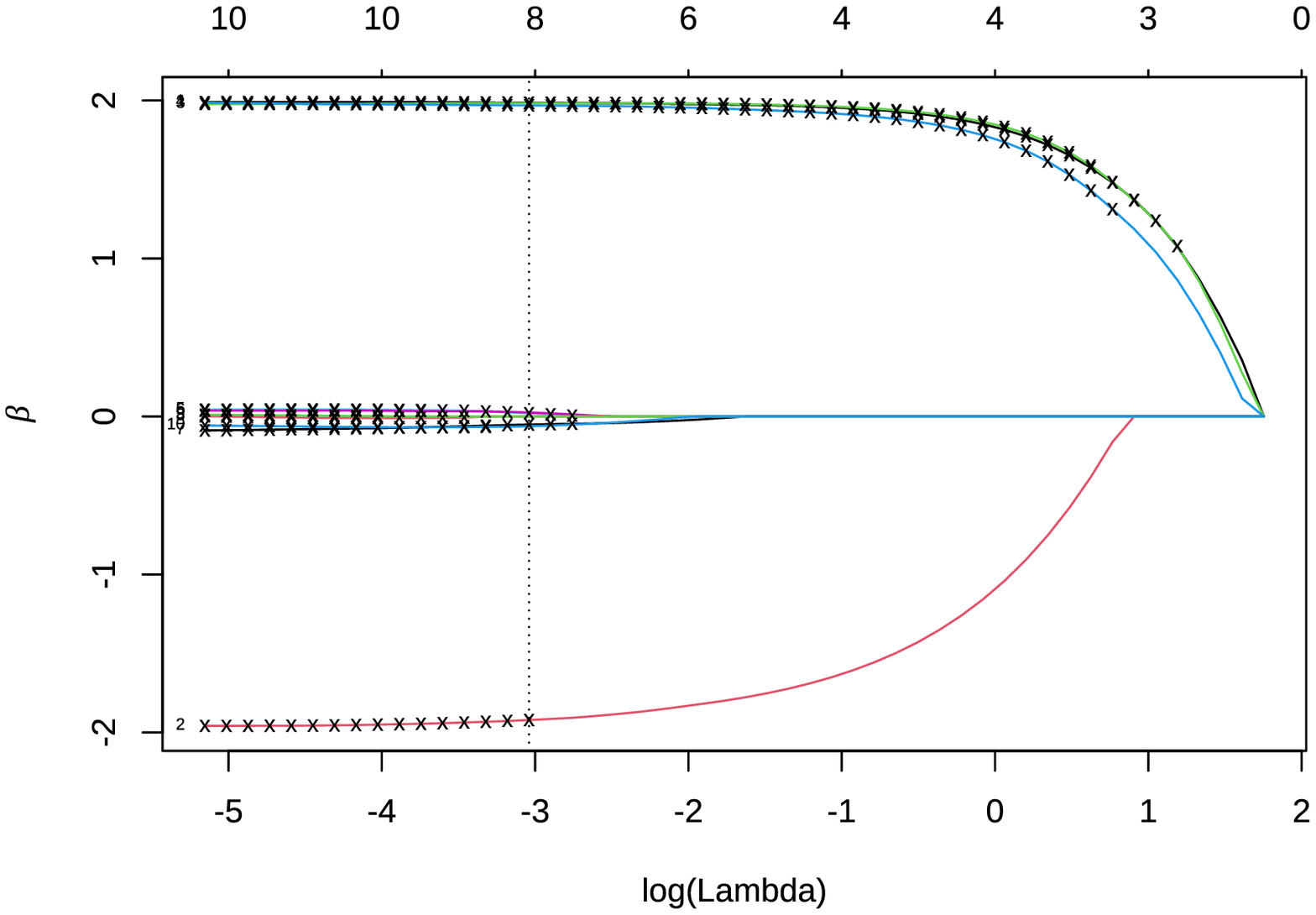}
 \caption{  }\label{pcoeff}
\end{subfigure}
\caption{Coefficient plots from the simulated experiment for $p=10$ for ADMM (a) and BCD (b). The numbers across the top of the validation plots represent the number of non-zero main effects at the given $\lambda$ and the vertical broken lines show the selected $\lambda$ from cross-validation. The “X” symbol indicates that a modifying term $Z\bm\theta_j$ has entered the model i.e the presence of an interaction effect.}\label{plot2}
\end{figure}

\begin{figure}[ht]
\begin{subfigure}{.33\textwidth}
  \centering
  \includegraphics[width=\textwidth]{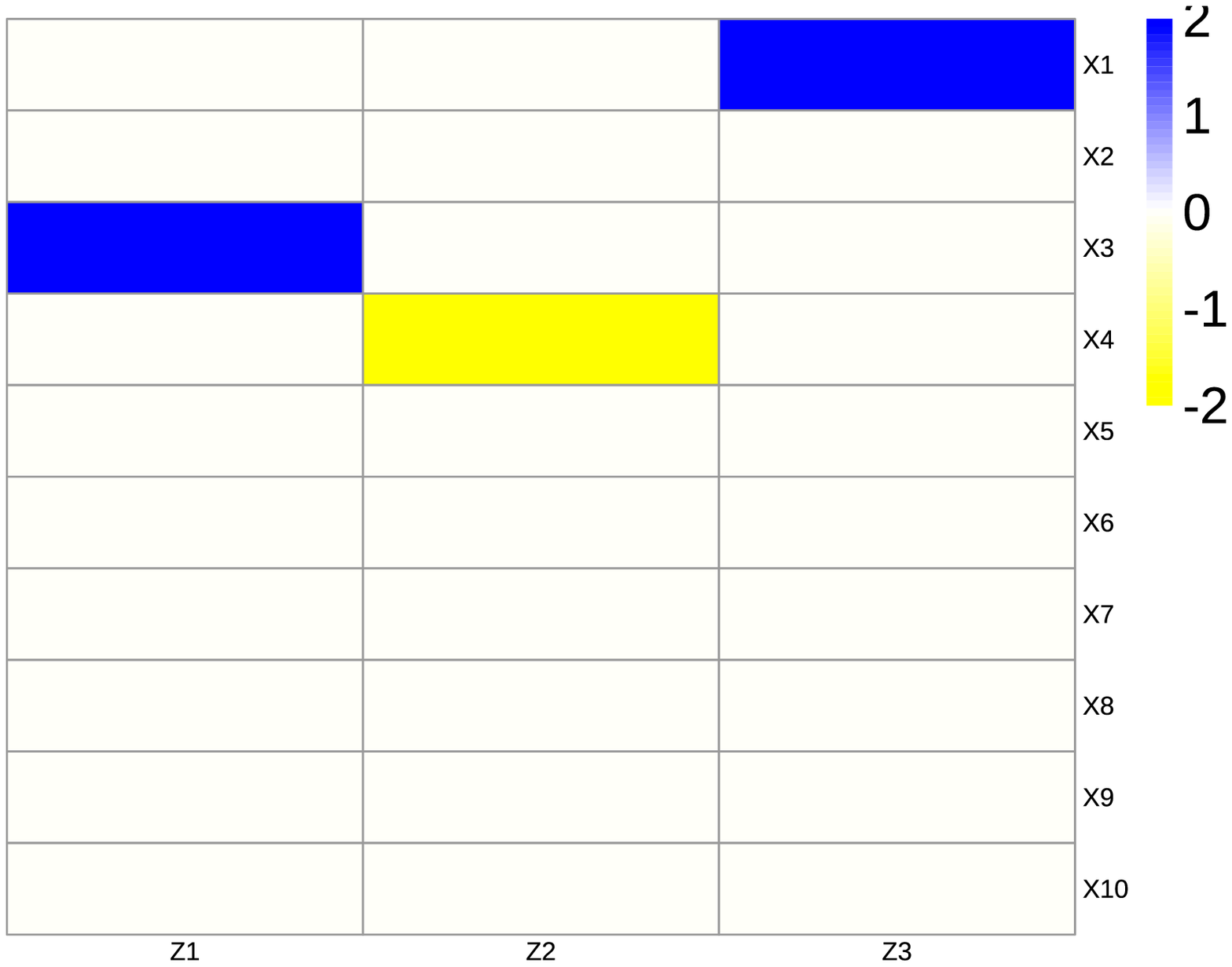}
 \caption{}\label{interplot}
\end{subfigure}
\begin{subfigure}{.33\textwidth}
  \centering
  \includegraphics[width=\textwidth]{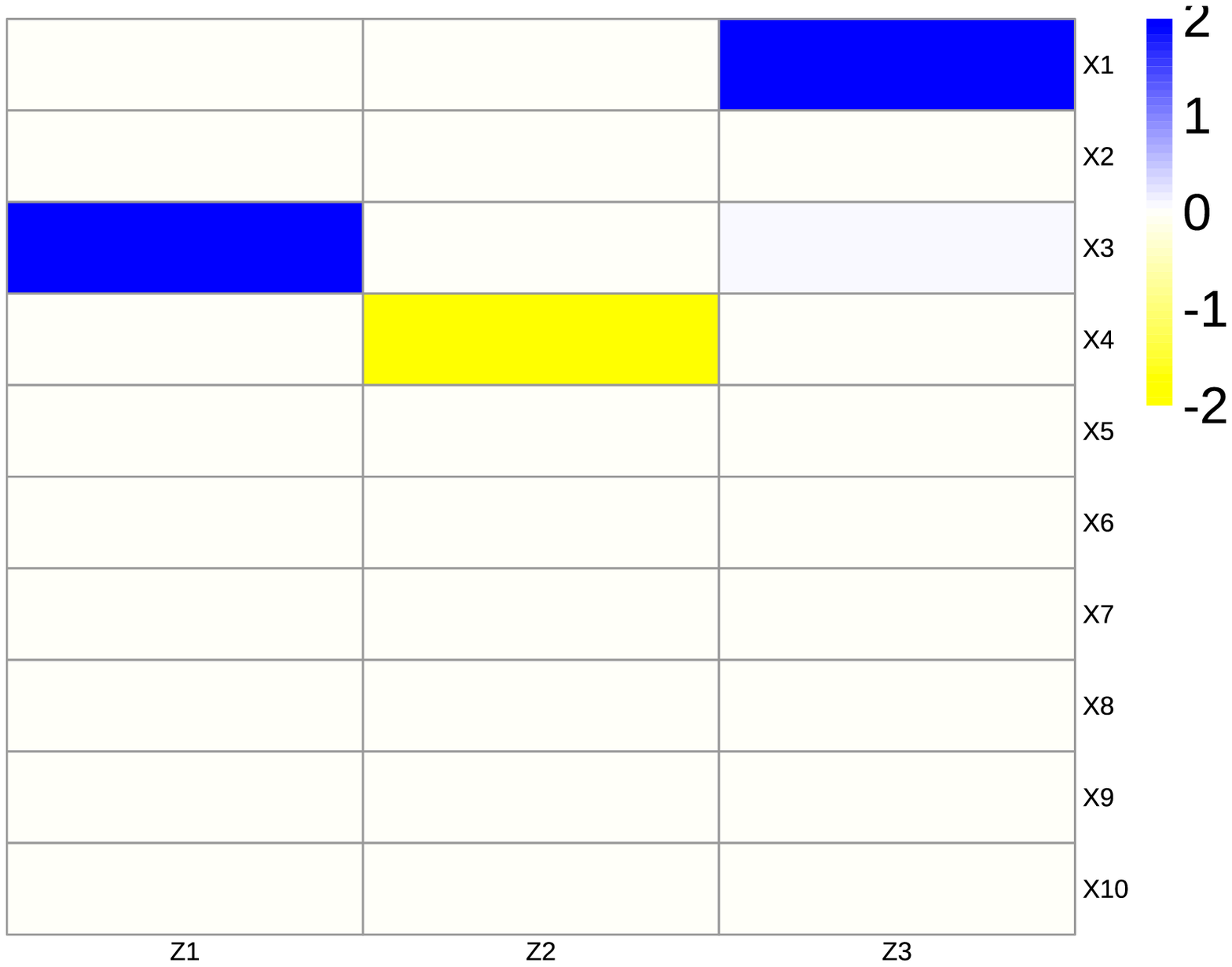}
 \caption{ }\label{interplot1}
\end{subfigure}
\begin{subfigure}{.33\textwidth}
  \centering
  \includegraphics[width=\textwidth]{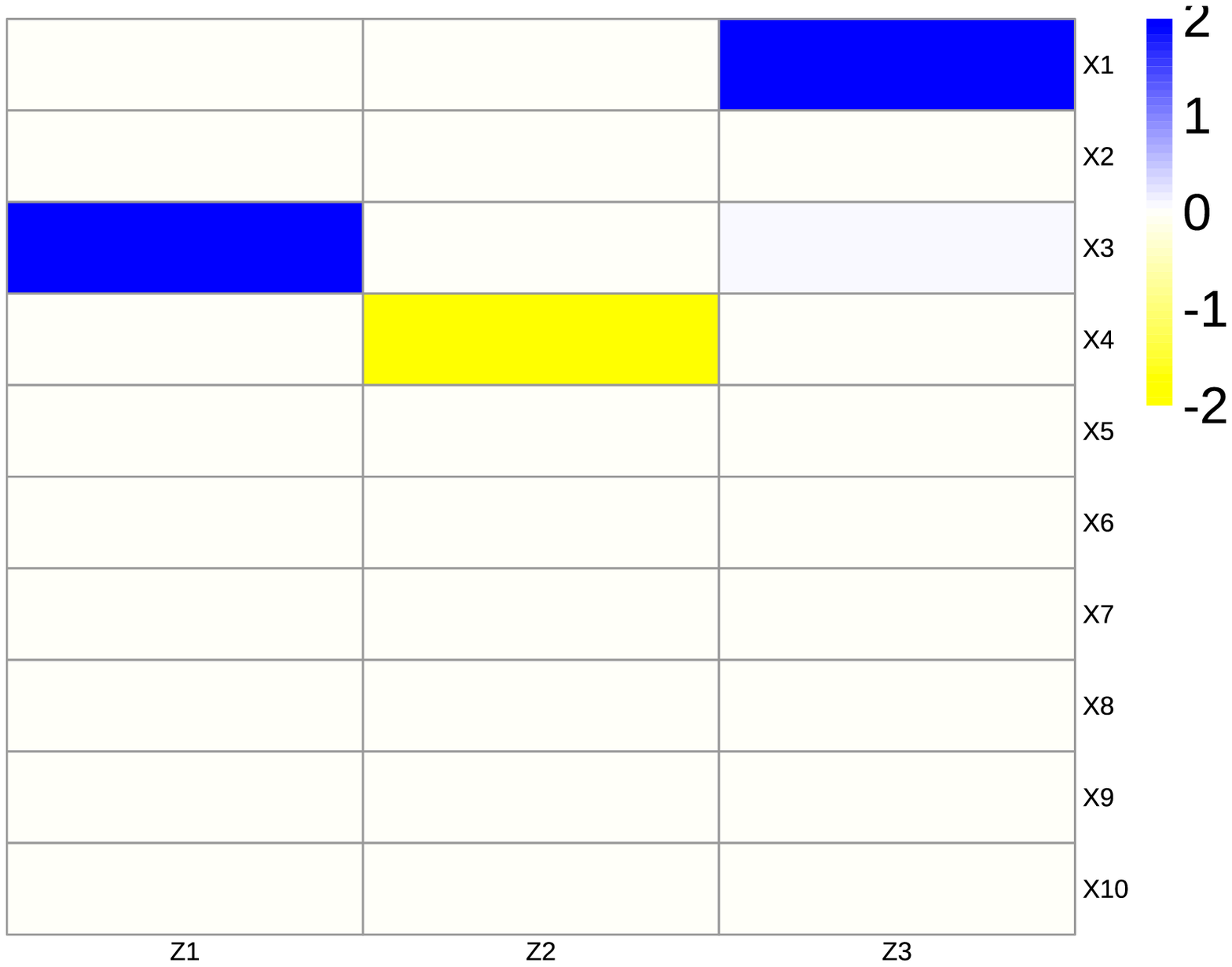}
 \caption{ }\label{interplot2}
\end{subfigure}
\caption{Matrix plot for the interaction coefficients $\theta_{jk}$ for $X_j$ and $Z_k$ from the simulated experiment for $p=10$ for the simulated values (a), ADMM estimates (b), and BCD estimates (c). The x-axis represents the modifying variables $Z_1,\ldots, Z_K$ and the y-axis represents the $X_1,\ldots,X_p$ to show the interactions $X_{j}\circ Z$.}\label{plot22}
\end{figure}

\begin{figure}[ht]
\begin{subfigure}{.5\textwidth}
  \centering
  \includegraphics[width=\textwidth]{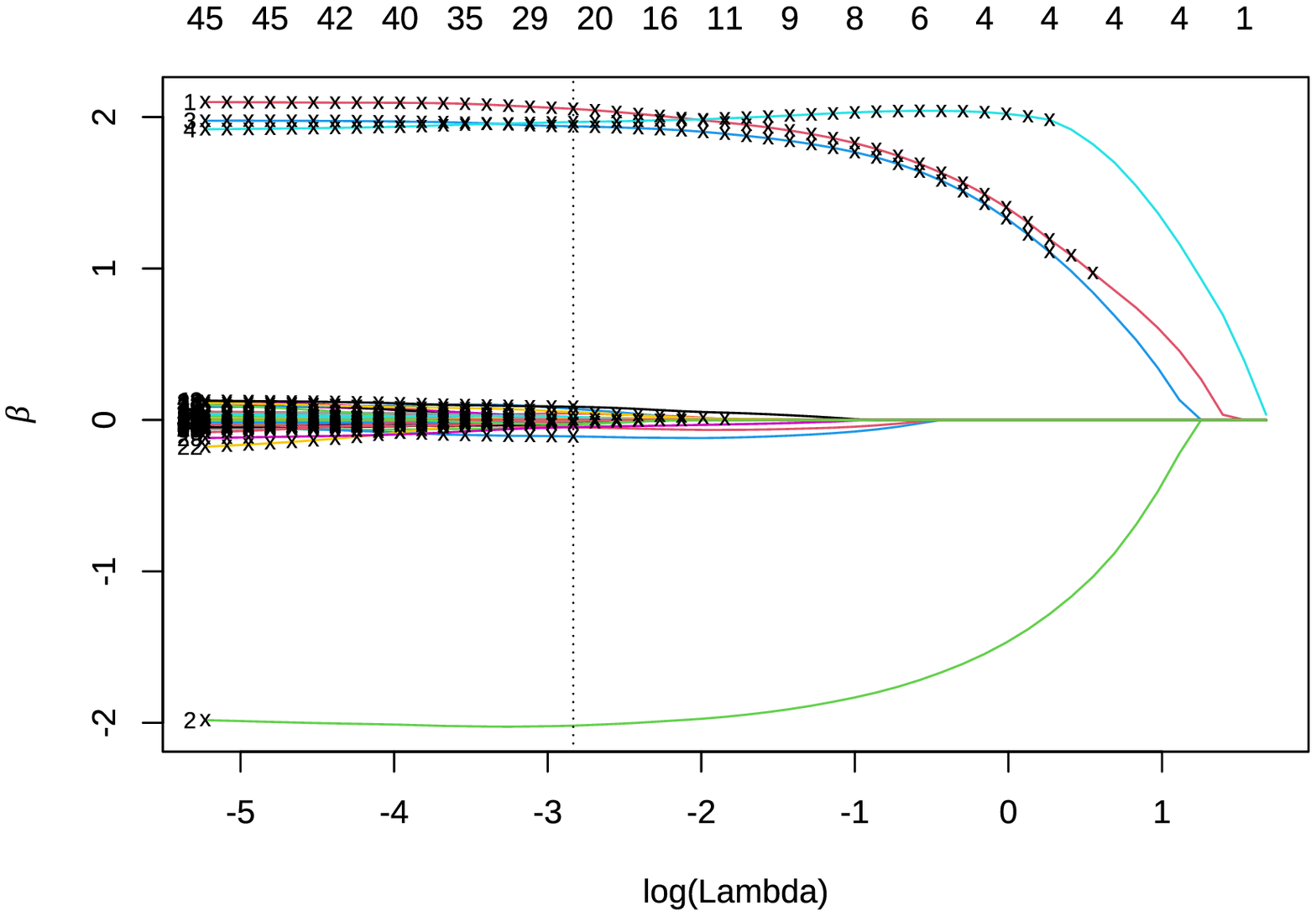}
 \caption{  }\label{adcoeff_500}
\end{subfigure}%
\begin{subfigure}{.5\textwidth}
  \centering
  \includegraphics[width=\textwidth]{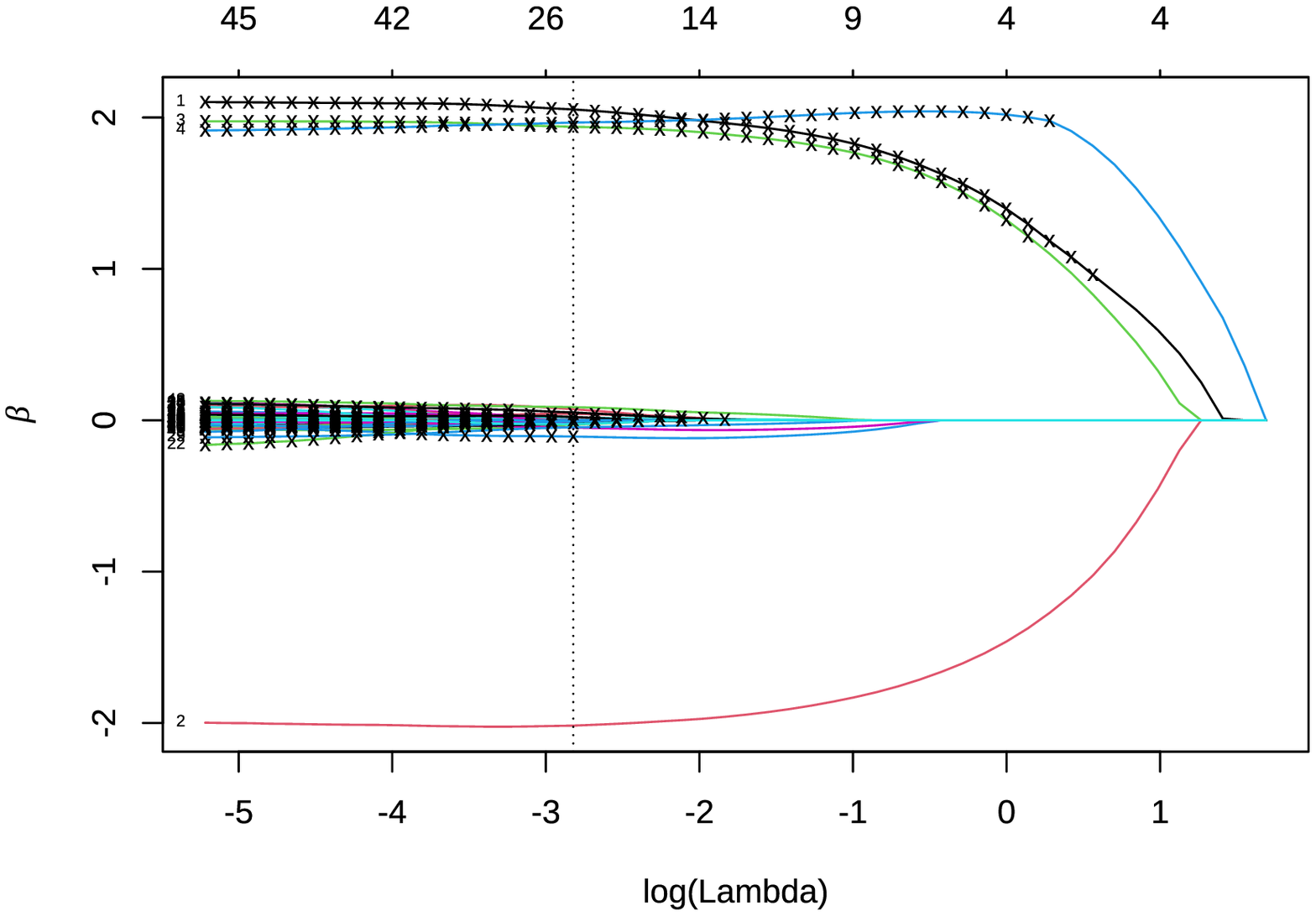}
 \caption{  }\label{pcoeff_500}
\end{subfigure}
\caption{Coefficient plots from the simulated experiment for $p=50$ for ADMM (a) and BCD (b). The numbers across the top of the validation plots represent the number of non-zero main effects at the given $\lambda$ and the vertical broken lines show the selected $\lambda$ from cross-validation. The “X” symbol indicates that a modifying term $Z\bm\theta_j$ has entered the model, i.e. the presence of an interaction effect.}\label{plot2_500}
\end{figure}

\begin{figure}[ht]
\begin{subfigure}{.33\textwidth}
  \centering
  \includegraphics[width=\textwidth]{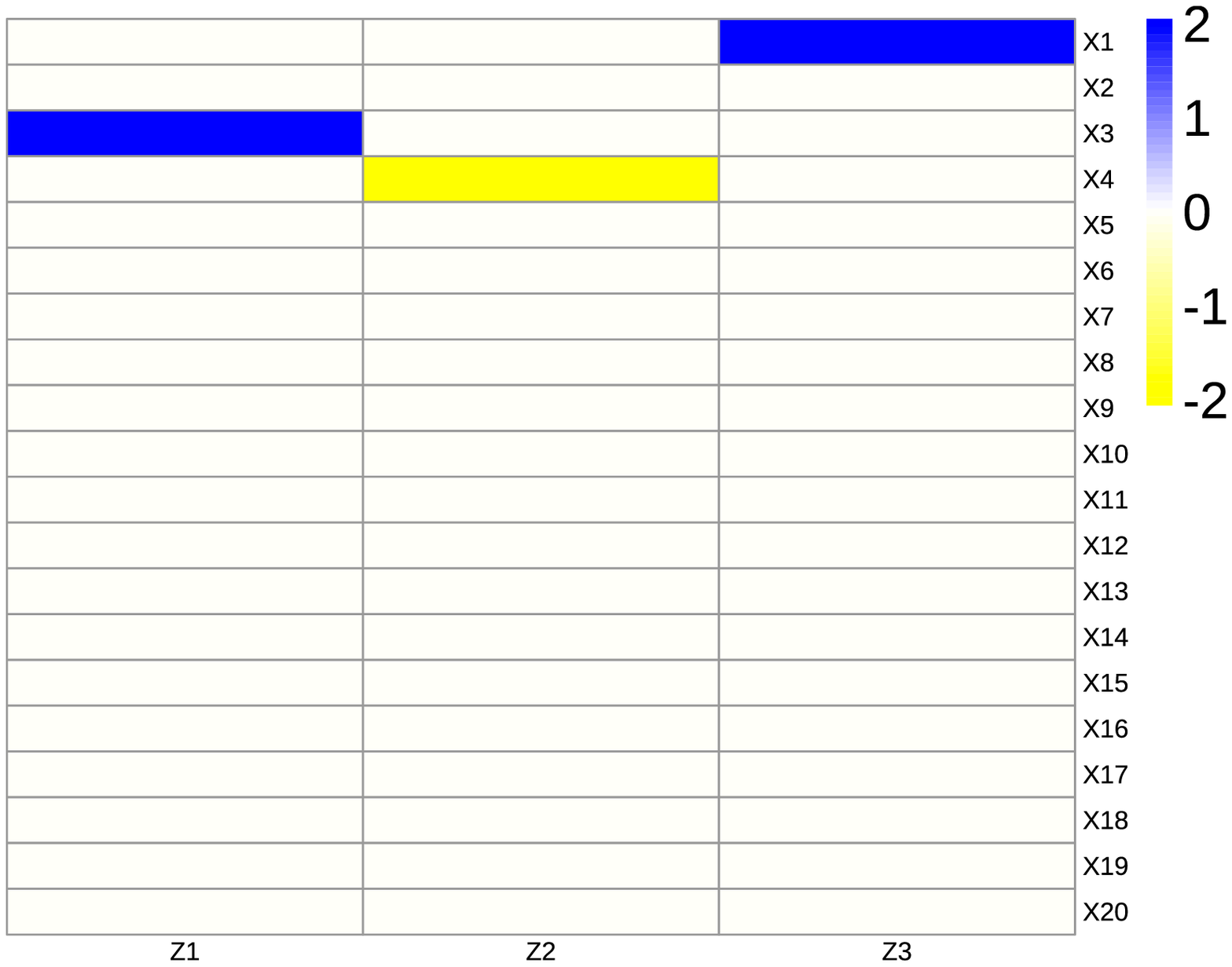}
 \caption{ }\label{interplot_500}
\end{subfigure}
\begin{subfigure}{.33\textwidth}
  \centering
  \includegraphics[width=\textwidth]{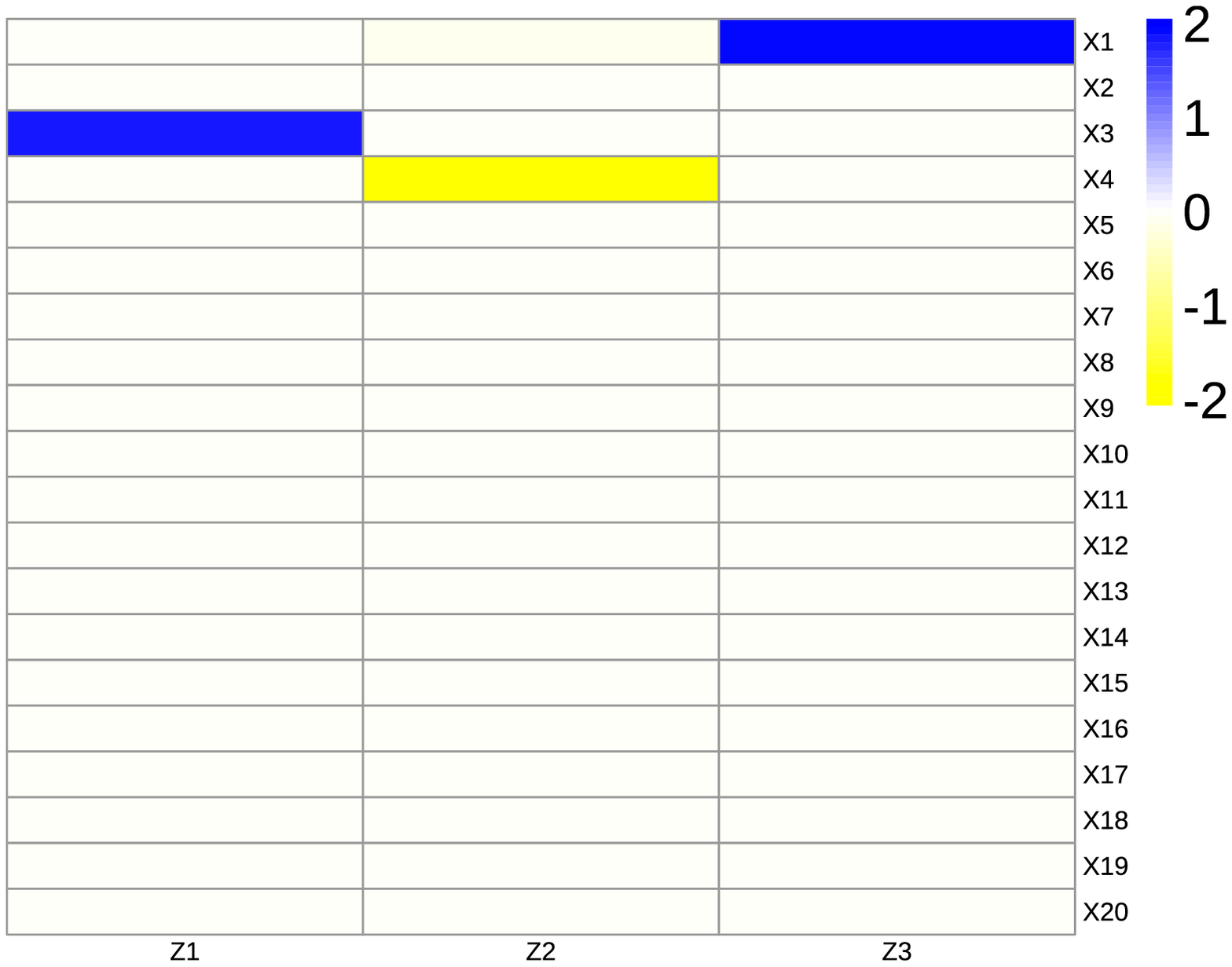}
 \caption{ }\label{interplot1_500}
\end{subfigure}
\begin{subfigure}{.33\textwidth}
  \centering
  \includegraphics[width=\textwidth]{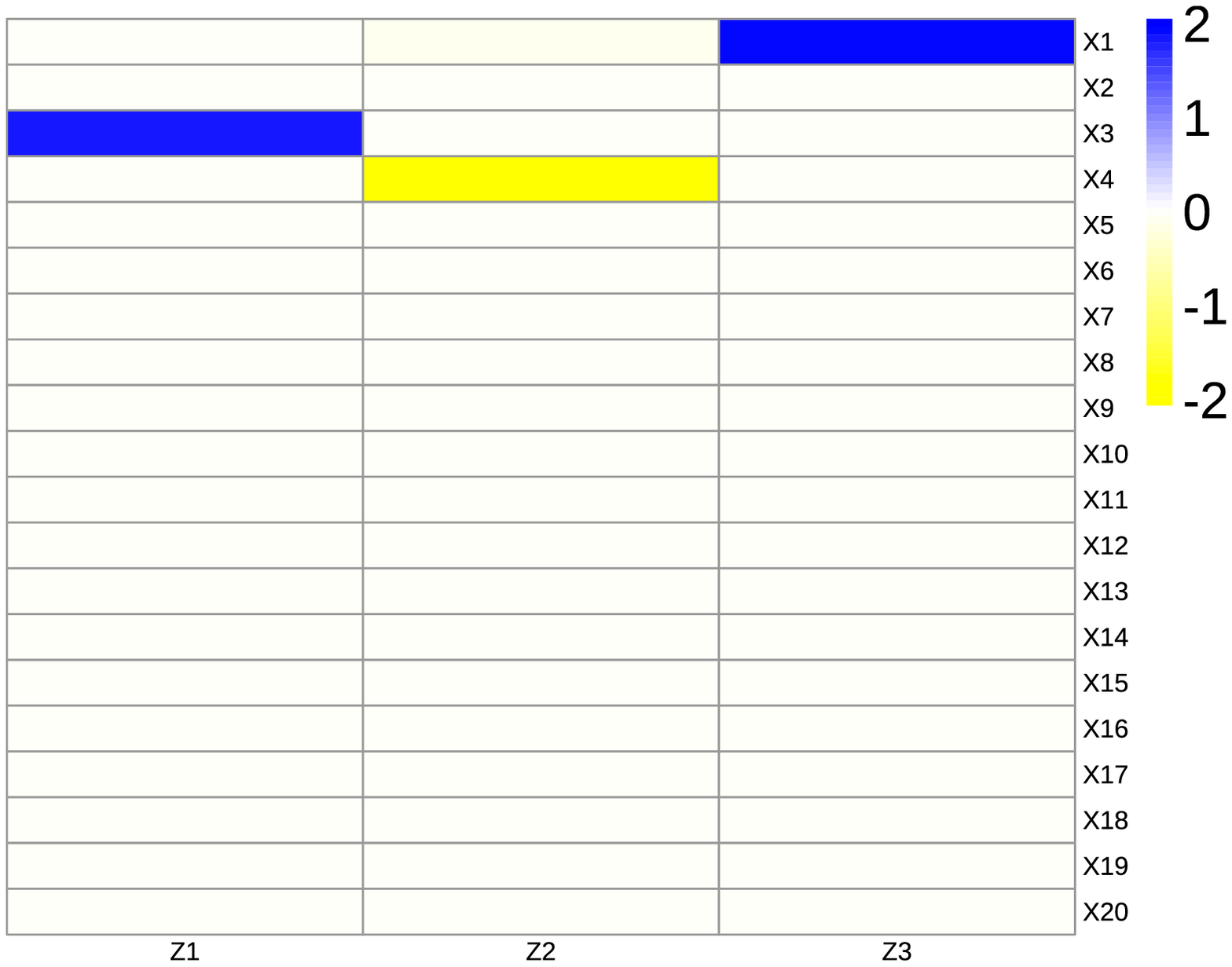}
 \caption{}\label{interplot2_500}
\end{subfigure}
\caption{Matrix plot for main and interaction coefficients from the simulated experiment for $p=50$ for the simulated values (a), ADMM estimates (b), and BCD estimates (c). The x-axis represents the modifying variables $Z_1,\ldots, Z_K$ and the y-axis represents the $X_1,\ldots,X_p$ to show the interactions $X_{j}\circ Z$.}\label{plot221}
\end{figure}

Figures \eqref{plot2} and \eqref{plot2_500} show the  results for the selected coefficients along the $\lambda$ path for ADMM, and the BCD for $p=10$ and $p=50$, respectively, after implementing them on the model \eqref{sim_model1}. The figures \eqref{plot22}  and \eqref{plot221}  show the matrix plot of the observed and the predicted coefficients of the interaction effects.  We can see that our proposed algorithm is able to identify  the true non-zero main effects as well as the non-zero interaction effects for model \eqref{sim_model1} starting from a position where all $\hat{\beta}_{j}$s are zero.

\clearpage

\subsection{Simulation of a multi-response problem}
To see the performance of the proposed method for the multi-response problem, we simulated two different  datasets; one with a weak hierarchical structure within the response variables and the other with a strong hierarchical structure. Both datasets were allowed to have both main effects $X$ and interactions with a modifying variable $Z$.   

\subsubsection{Simulation 1}
For the first simulation, we generated a multi-response problem with $D=6$ responses, $N=100$ observations, $p=500$ main effects and $K=4$ modifying variables. Both $X$ and $Z$ were generated as standard Gaussian independent predictors. We allowed each response to have five non-zero main coefficients and also allowed four $X_j$s in each response to have interactions with four $Z_k$s. The error term, $\mathbb{E}_d\sim \mathbf{N}(0,1)$ was added to each response $d=1,\ldots,D$ and we generated the responses from 
\begin{equation}
    Y=XB+\sum_d\underset{j}{\sum}(X_{j}\circ Z)\bm\theta_{jd}+\mathbb{E}.
\end{equation} 
We averaged the results over 10 simulations and compared our work with that of the tree lasso, which has been made available in R in the package `mixlasso' version 0.1 available on Github (https://github.com/zhizuio/mixlasso) and the pliable lasso \citep{tib} package (‘pliable’ version 1.1.1) also available in CRAN archive (https://cran.r-project.org/src/contrib/Archive/pliable/). To allow the tree lasso model to include the modifying variable, we allowed the modifier $Z$ to be included without any penalty. This option has been made available in the mixlasso R package by selecting \texttt{method="tree-lasso"} and \texttt{num.nonpen=K}. 

\begin{figure}[ht]
\begin{subfigure}{.5\textwidth}
  \centering
  \includegraphics[width=\textwidth]{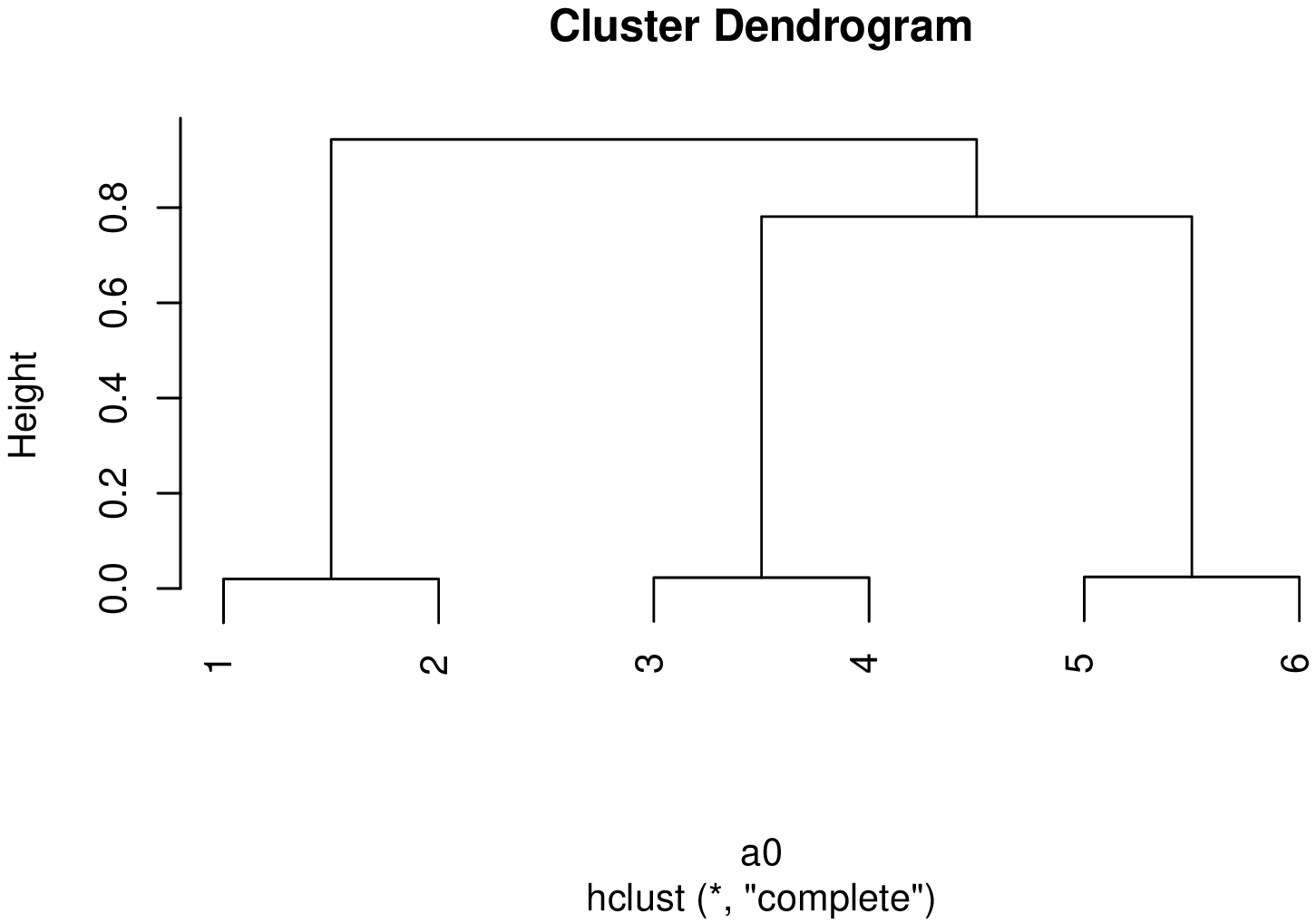}
  \caption{ }\label{tree_sim1}
\end{subfigure}
\begin{subfigure}{.5\textwidth}
  \centering
  \includegraphics[width=.9\textwidth]{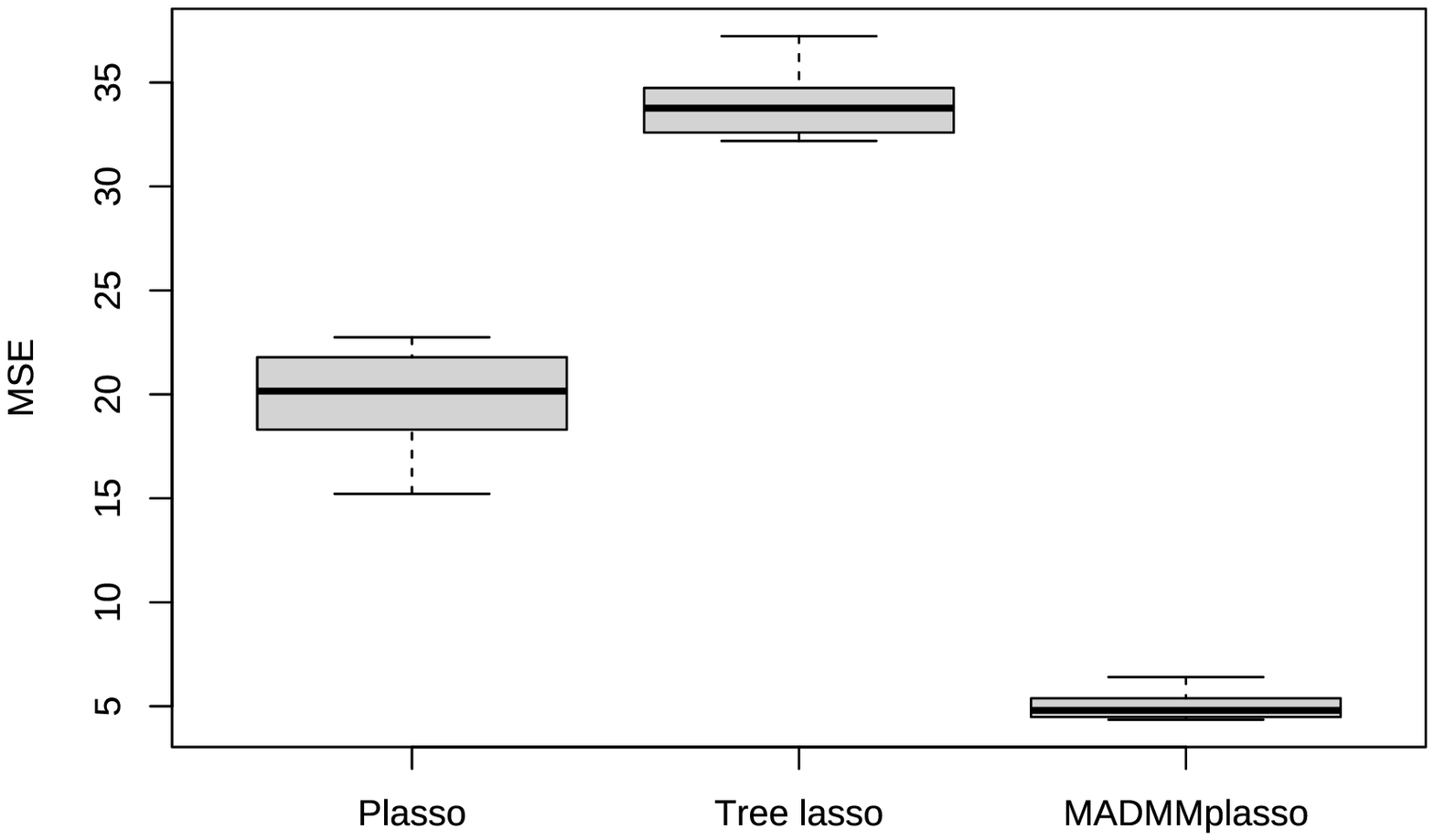}
 \caption{ }\label{Box1}
\end{subfigure}
\caption{The general structure of multi-response simulation 1 and the box plot from the mean squared errors (MSE) on the test data. The dendrogram (a) shows the hierarchical structure of the multivariate responses generated by $\beta$, as estimated with complete-linkage hierarchical clustering using the Euclidean distance. Below each node of the dendrogram is a number indicating the response variable associated with the node.  The MSE on the test data (b) was averaged across the 10 simulations.  }\label{tree_sim1_col}
\end{figure}

The structure of the response variables $Y$ is shown in figure \eqref{tree_sim1}. The summary of the results is shown in table \eqref{table1}, where we summarize the sensitivity (the proportion of non-zero coefficients correctly estimated as non-zeros), specificity (the proportion of zero-coefficients correctly estimated as zeros), the total number of the estimated non-zero coefficients (out of $p=500\times D=6$) and the test error measured as the mean squared error on a separate test data with $N=500$, $p=500$ and $K=4$. It can be seen that our proposed model performs best in this case in terms of the test error (from figure \eqref{Box1}). In this setting with weak hierarchical structure in the response, the pliable lasso model performed better than tree lasso, indicating that the correctional structure in the responses is not so important. By looking at the averaged absolute errors of estimated coefficients ($1/Dp\lVert\hat{\beta}-\beta\rVert_1$) in table \eqref{table1}, it can be seen that the proposed model also performs best in terms of the accuracy of coefficients matrix estimation and variable selection for this particular case. 

\begin{table}[ht]
\small 
  \centering
    \caption{Results from the multi-response simulation 1 with weak hierarchical structure in the response.}
    \label{table1}
    \begin{threeparttable}
    
    \begin{tabular}{|l l l l l l|} 
    \hline
     \textbf{Model} & $(1/D p)\lVert \hat{\beta}-\beta\rVert_1$ & Sensitivity\tnote{1} & Specificity\tnote{2} & Non-zero\tnote{3} & Test error (SD)\tnote{4} \\
      \hline
     Plasso & 0.021 & 1 & 0.763 & 733 & 19.693 (2.408) \\
     Tree lasso & 0.066 & 1 & 0.142 & 2577 & 34.045 (1.802) \\
MADMMplasso & 0.006 & 1 & 0.991 & 237 & 5.050 (0.681) \\     
     \hline
     
\end{tabular}
    \begin{tablenotes}
    \item[1] Sensitivity is the proportion of non-zero coefficients estimated as non-zeros. 
    \item[2] Specificity is the proportion of zero-coefficients estimated as zeros.
    \item[3] The total number of non-zero coefficients in the model. We counted  the coefficients with at least two non-zero values across the 10 simulations. 
    
    Number of non-zero coefficients $=\sum_{j=1}^p\sum_{d=1}^D\{(\sum_{r=1}^{10} \bm1_{\{\beta_{jd}^r\neq 0\}})\geq 2\}$. Note that the selection is out of $p\times D=3000$ features in total. 
    \item[4] The MSE on an independent test dataset. We include the standard deviation (SD) across the 10 simulations.
  \end{tablenotes}
    \end{threeparttable}

\end{table}

\begin{figure}[ht]
  \centering
  \includegraphics[width=\textwidth]{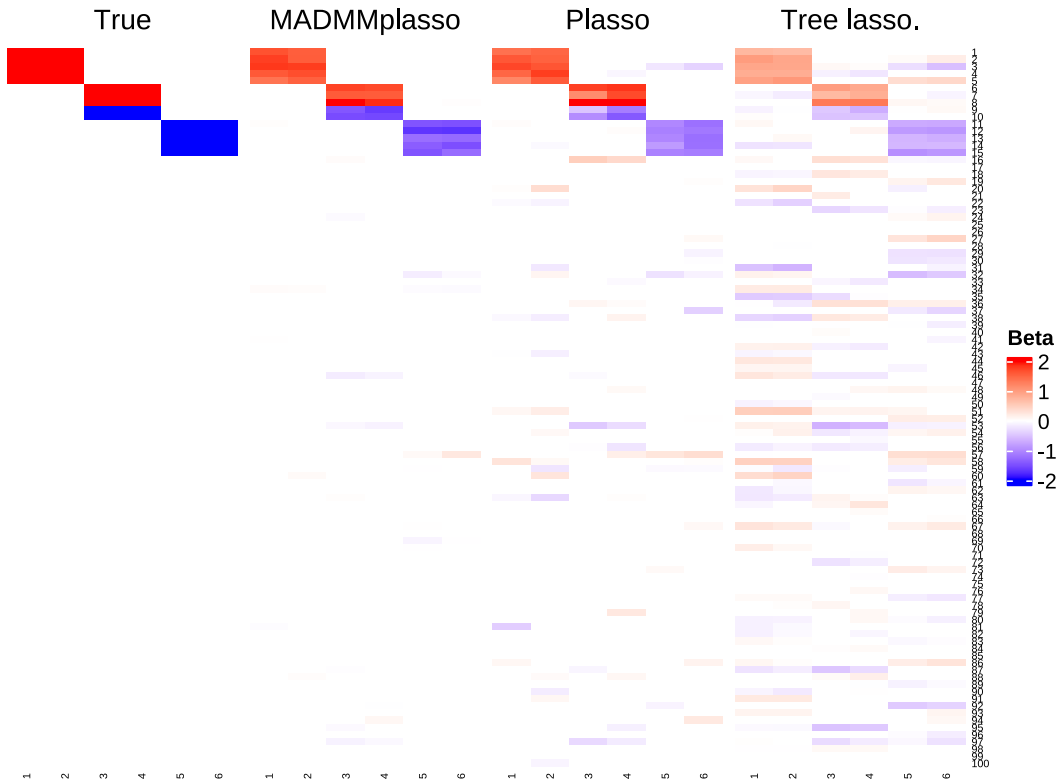}
\caption{The general structure of the multi-response simulation 1. Starting from left, the simulated (``true'') coefficients, MADMMplasso estimates, plasso estimates, and tree lasso estimates. For the true coefficients, we allowed each response to have five non-zero coefficients as indicated by the red and blue blocks. Below each block is a number indicating the response variable associated with the block. Some responses were allowed to have non-zero coefficients for the same variables so that we could obtain some correlations among the responses.   }\label{heat_sim1_col}
\end{figure}

Figure \eqref{heat_sim1_col} shows the structure of the main effect coefficients ($\beta$) and how they correlate across the responses. It can be seen that our proposed model is able to recover the true relevant covariates for correlated responses significantly better than other methods. The plasso model seem not to handle the groups well since it not able to borrow statistical strength across different responses. Even though the tree lasso does well in recovering the groups, it seem to have a lot of false positives than our proposed model.

\subsubsection{Simulation 2}
For the second simulation in this section, we generated a multi-response problem with $D=24$ responses (imagined to be responses of cancer cells to drugs), $N=100$ observations (e.g. different cancer cell lines), $p=150$ or $p=500$ (e.g. gene expression variables characterizing the different cancer cell lines) and $K=4$ (e.g. cancer tissue type). The idea is to simulate a drug sensitivity response model that includes interaction effects to allow the cancer tissue types to modify the effects of gene expression features.  $X$  was generated as standard Gaussian independent predictors to represent the gene expression features and the $Z$ was was generated using the binomial distribution to represent the presence or absence of a certain tissue. The interaction in this scenario was to allow for inclusion of tissue effect, where we have assumed that certain drugs work best on specific tissues. The error term, $\mathbb{E}_d\sim \mathbf{N}(0,1)$ was added to each response $d=1,\ldots,D$ and we generated the drug sensitivity using the model 
\begin{equation}
    Y=XB+\sum_d\sum_j(X_{j}\circ Z)\bm\theta_{jd}+\mathbb{E}.
\end{equation} We generated $X$ by following the example in \cite{zhao2020structured}. However, since we have only one source of data in $X$, we decided to only use the first part of their matrix. To do this, we generated $X$ from the multivariate normal distribution with zero  mean and non-diagonal $p\times p$ covariance matrix $\sum$. The covariance matrix $\sum$ has $p/b(\sigma)$ as the covariance of features and the remaining part being set to zero. We set $\sigma =0.4$, $b=10$, and the variance of each feature is one. We averaged the results over 10 simulations and compared our work with that of the tree lasso and the pliable lasso. To allow the tree lasso model to include the modifying variable, we allowed the modifier $Z$ to be included without any penalty. This option has been made available in the mixlasso R package selecting \texttt{method="tree-lasso"} and \texttt{num.nonpen=K}. 

\begin{figure}[ht]
\begin{subfigure}{.33\textwidth}
  \centering
  \includegraphics[width=\textwidth]{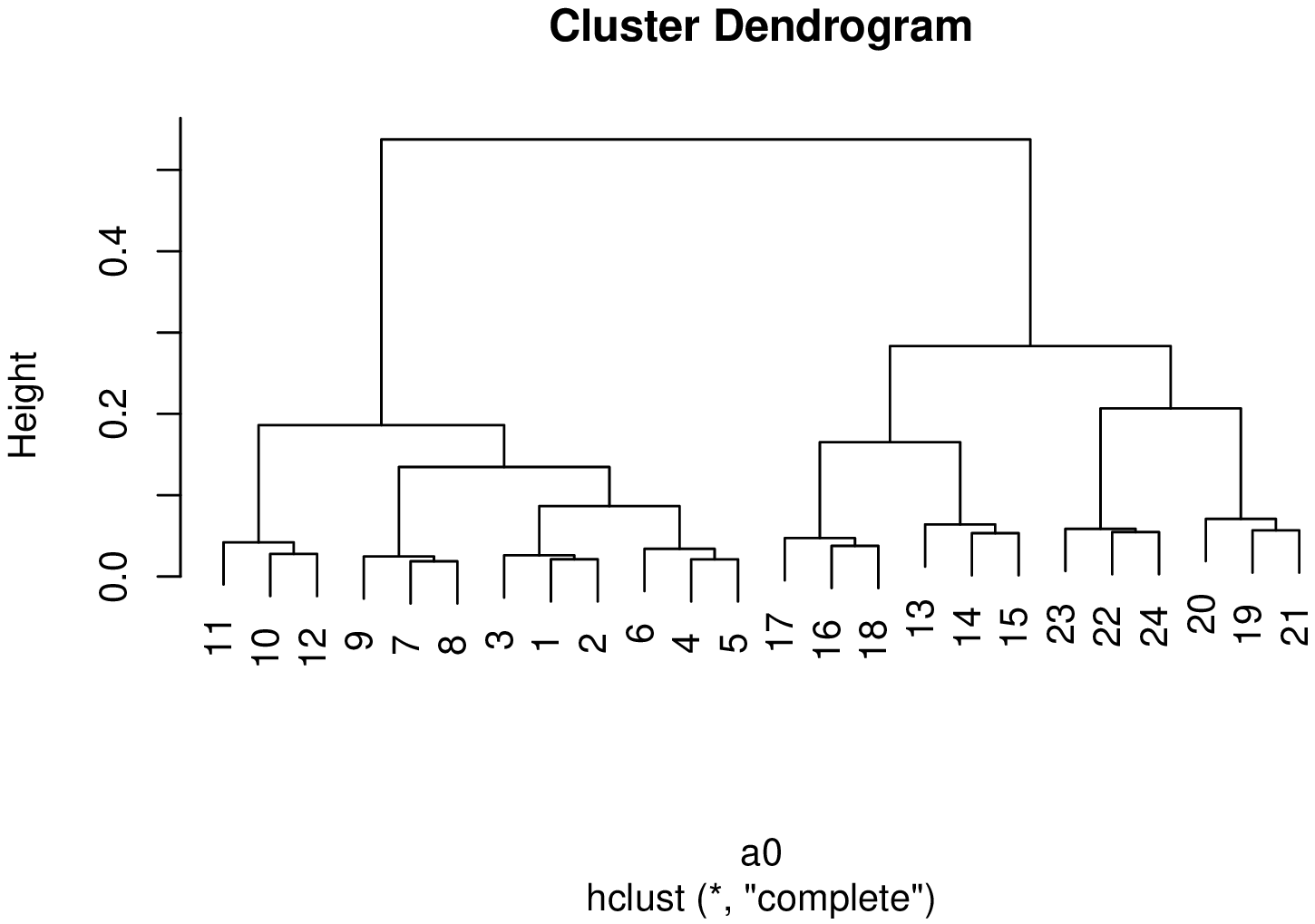}
 \caption{ }\label{tree2}
\end{subfigure}
\begin{subfigure}{.33\textwidth}
  \centering
  \includegraphics[width=\textwidth]{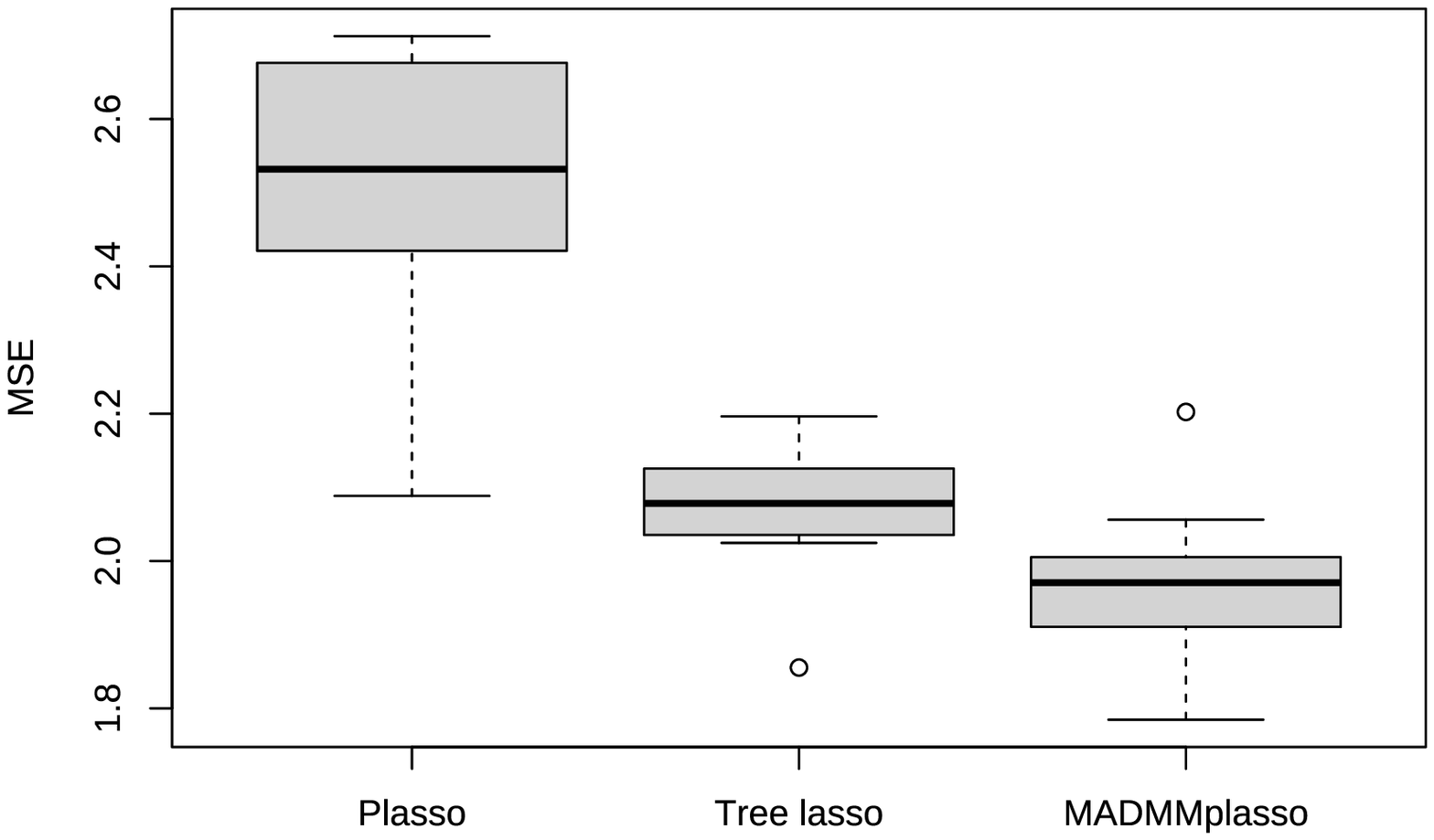}
 \caption{ }\label{Box2}
\end{subfigure}
\begin{subfigure}{.33\textwidth}
  \centering
  \includegraphics[width=\textwidth]{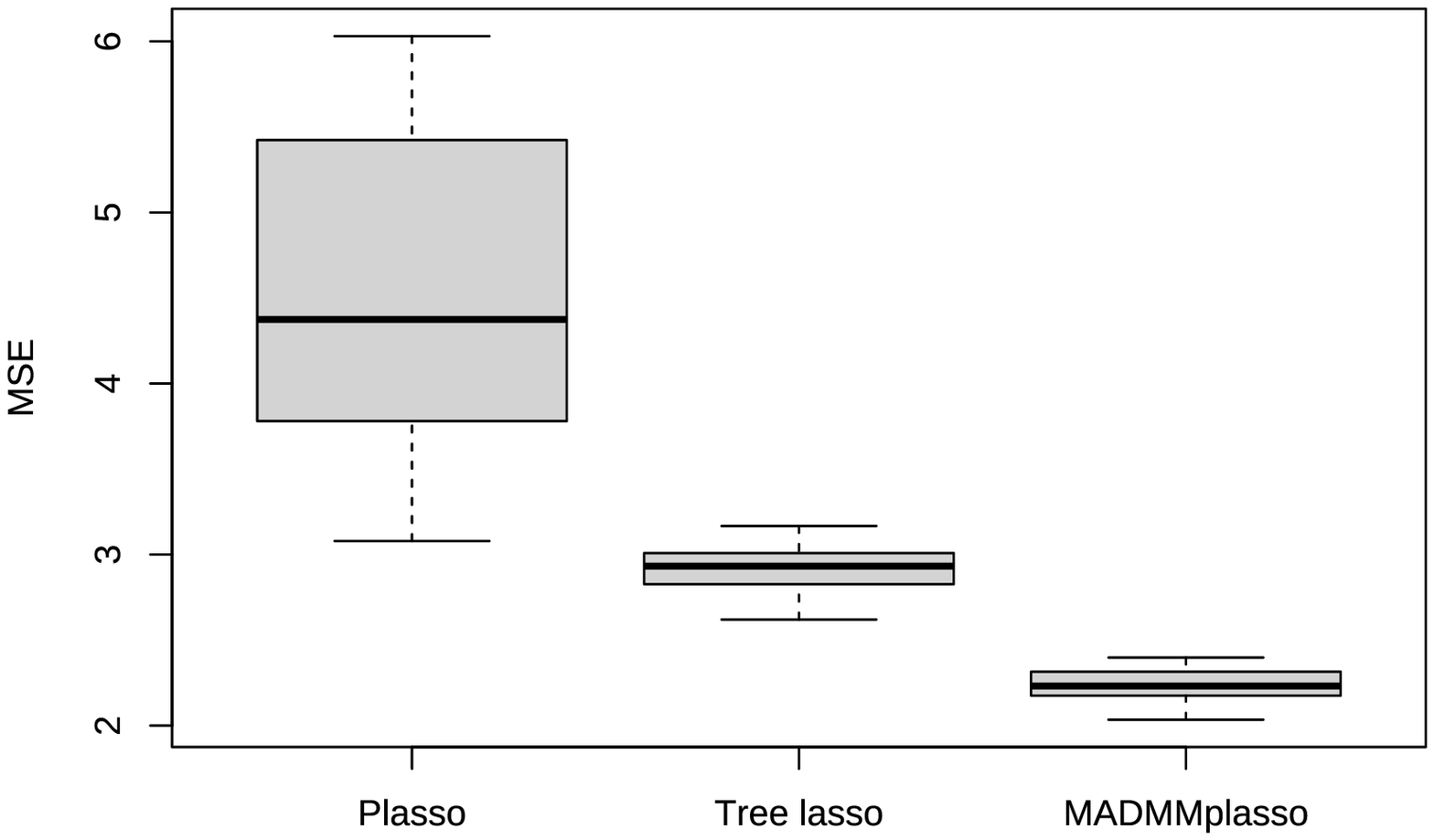}
 \caption{ }\label{Box3}
\end{subfigure}
\caption{The general structure of the multi-response simulation 2 and the box plots from the mean squared error (MSE) on the test data. The dendrogram (a) shows the hierarchical structure of the multivariate responses generated by $\beta$, as estimated with complete-linkage hierarchical clustering using the Euclidean distance. Below each node of the dendrogram is a number indicating the response variable associated with the node. The MSE on the test data for $p=150$ (b), and $p=500$ (c) were averaged across the 10 simulations. }\label{tree_sim22_col}
\end{figure}

The structure of the response variables $Y$ is shown in figure \eqref{tree2}. The results are shown in table \eqref{table2}, where we summarize the sensitivity, specificity, the total number of non-zero coefficients (out of $p\times D$) and the test error measured as the mean squared error on a separate test dataset. Our proposed model performs best in this case in terms of the test error (in figures \eqref{Box2} and \eqref{Box3}). In this setting with strong hierarchical structure in the responses, the tree lasso performs better than the pliable lasso, indicating the importance of the groups within the responses. 

By looking at the the averaged absolute errors of estimated coefficients ($1/Dp\lVert\hat{\beta}-\beta\rVert_1$) in table \eqref{table2}, we see that the proposed model also performs best in terms of the accuracy of coefficients matrix estimation and variable selection for this particular case. 

\begin{table}[ht]
\small 
\centering
    \caption{Results from the multi-response simulation 2 with strong hierarchical structure in the responses. }
    \label{table2}
    \begin{threeparttable}
    \begin{tabular}{|l l l l l l|} 
    \hline
     \textbf{Model} & $(1/D p)\lVert \hat{\beta}-\beta\rVert_1$ & Sensitivity\tnote{1} & Specificity\tnote{2} & Non-zero\tnote{3} & Test error (SD)\tnote{4} \\
      \hline
      $p=150$ & & & & & \\
     Plasso & 0.034 & 1 & 0.446 & 2155 & 2.512 (0.181) \\
     Tree lasso & 0.036 & 1 & 0.345 & 2483 & 2.072 (0.095) \\
MADMMplasso & 0.0299 & 1 & 0.727 & 2014 & 1.972 (0.112) \\   
$p=500$ & & & & & \\
     Plasso & 0.014 & 0.994 & 0.814 & 2514 & 4.57 (1.038) \\
     Tree lasso & 0.023 & 1 & 0.360 & 7826 & 2.927 (0.163) \\
MADMMplasso & 0.010 & 1 & 0.912 & 1891 & 2.230 (0.116) \\    
     \hline
     
    \end{tabular}
    \begin{tablenotes}
    \item[1] Sensitivity is the proportion of non-zero coefficients estimated as non-zeros. 
    \item[2] Specificity is the proportion of zero-coefficients estimated as zeros.
    \item[3] The total number of non-zero coefficients in the model. We counted the coefficients with at least two non-zero values across the 10 simulations. 
   
    Number of non-zero coefficients $=\sum_{j=1}^p\sum_{d=1}^D\{(\sum_{r=1}^{10} \bm1_{\{\beta_{jd}^r\neq 0\}})\geq 2\}$. Note that the selection is out of $p\times D=3600$ (for $p=150$) or
    $12000$ (for $p=500$) features in total. 
    \item[4] The MSE on an independent test dataset. We included the standard deviation (SD) across the 10 simulations.
  \end{tablenotes}
    \end{threeparttable}
\end{table}

\begin{figure}[ht]
  \centering
  \includegraphics[width=\textwidth]{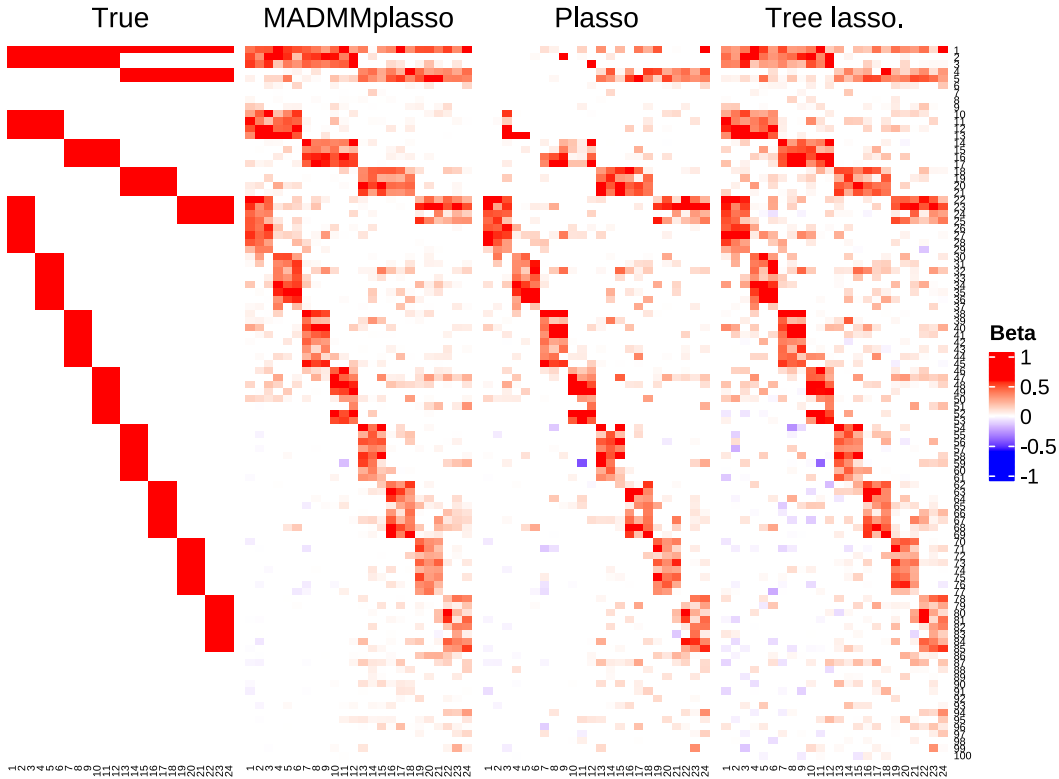}
\caption{The general structure of the multi-response simulation 2. Starting from left, the simulated (``true'') coefficients, MADMMplasso estimates, plasso estimates, and tree lasso estimates. The non-zero coefficients in the true data are represented by red blocks. Below each block is a number indicating the response variable associated with the block. Some responses were allowed to have non-zero coefficients for the same variables so that we could obtain some correlations among the responses.    }\label{heat_sim2150_col}
\end{figure}

Figure \eqref{heat_sim2150_col} shows the structure of the main effect coefficients ($\beta$) and how they correlate across the responses. It is clear that the tree lasso and our proposed method are able to recover the true relevant covariates for correlated responses significantly better than the pliable lasso even though the tree lasso seems to have rather more false positives (also see table \ref{table2}).

\section{Genomics of drug sensitivity in cancer (GDSC) data analysis}\label{section4}
\begin{figure}[ht]
\begin{subfigure}{.5\textwidth}
  \centering
  \includegraphics[width=\textwidth]{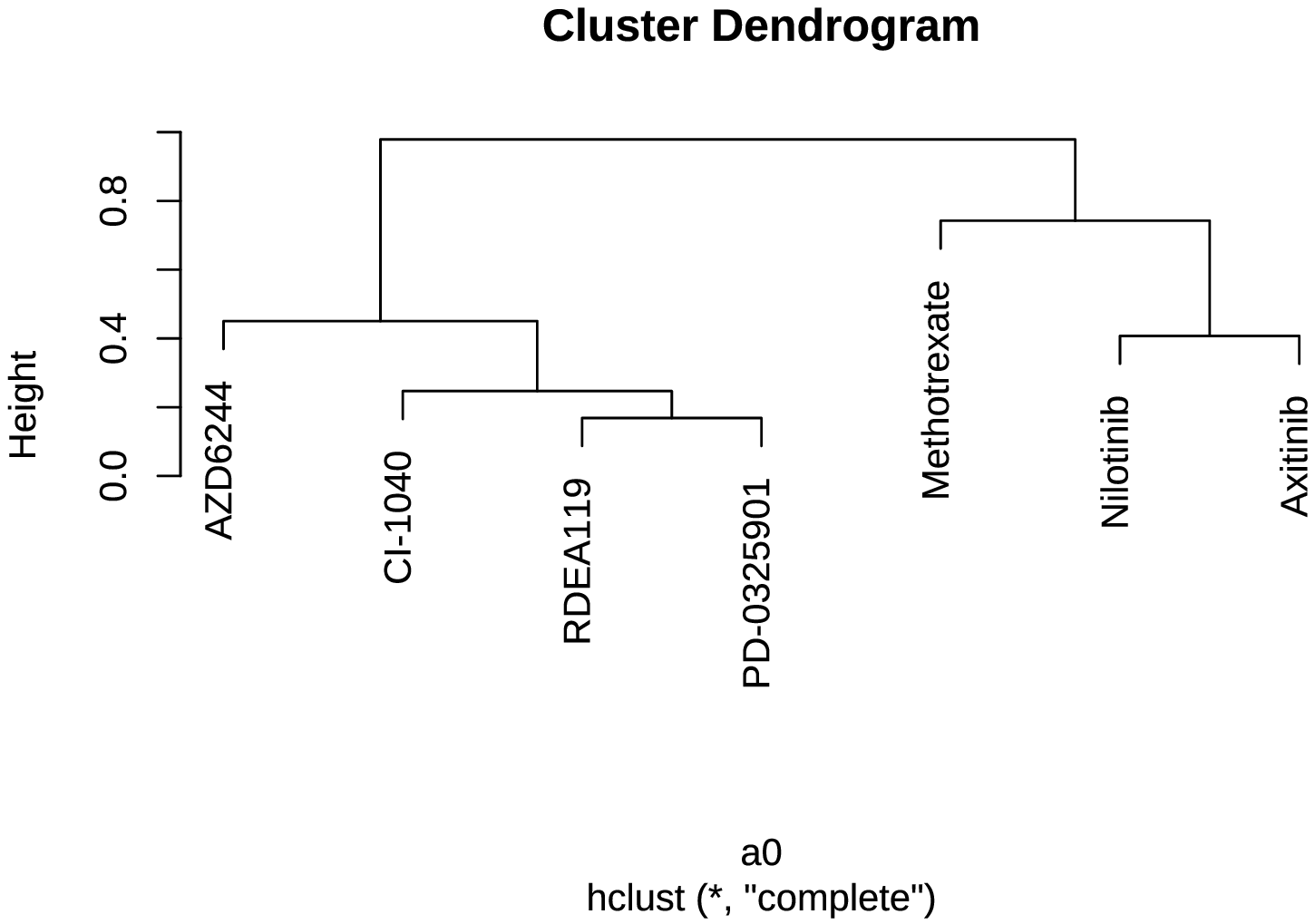}
 \caption{ }\label{tree3}
\end{subfigure}%
\begin{subfigure}{.5\textwidth}
  \centering
  \includegraphics[width=\textwidth]{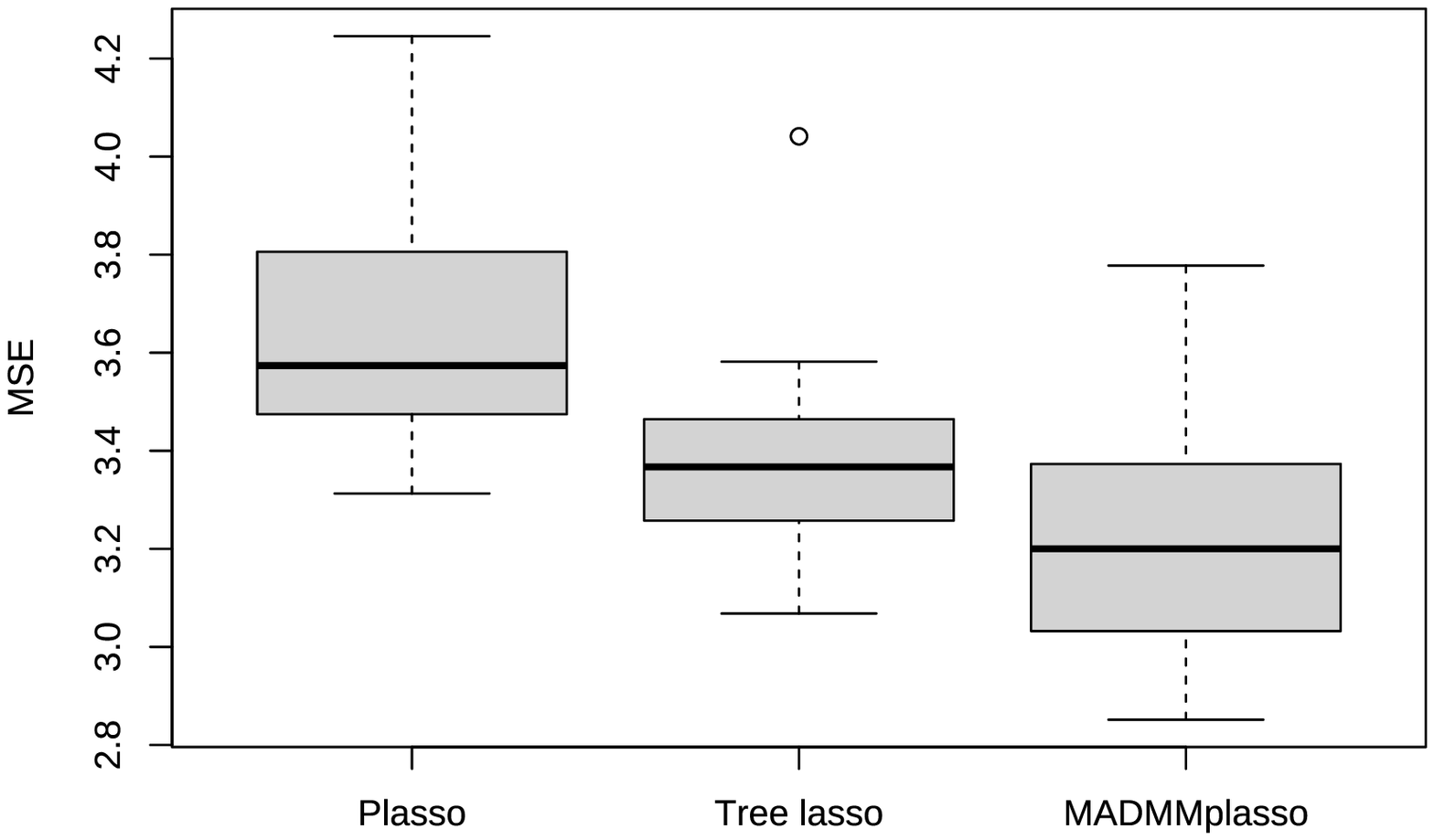}
 \caption{ }\label{Box4}
\end{subfigure}
\begin{subfigure}{.5\textwidth}
  \centering
  \includegraphics[width=\textwidth]{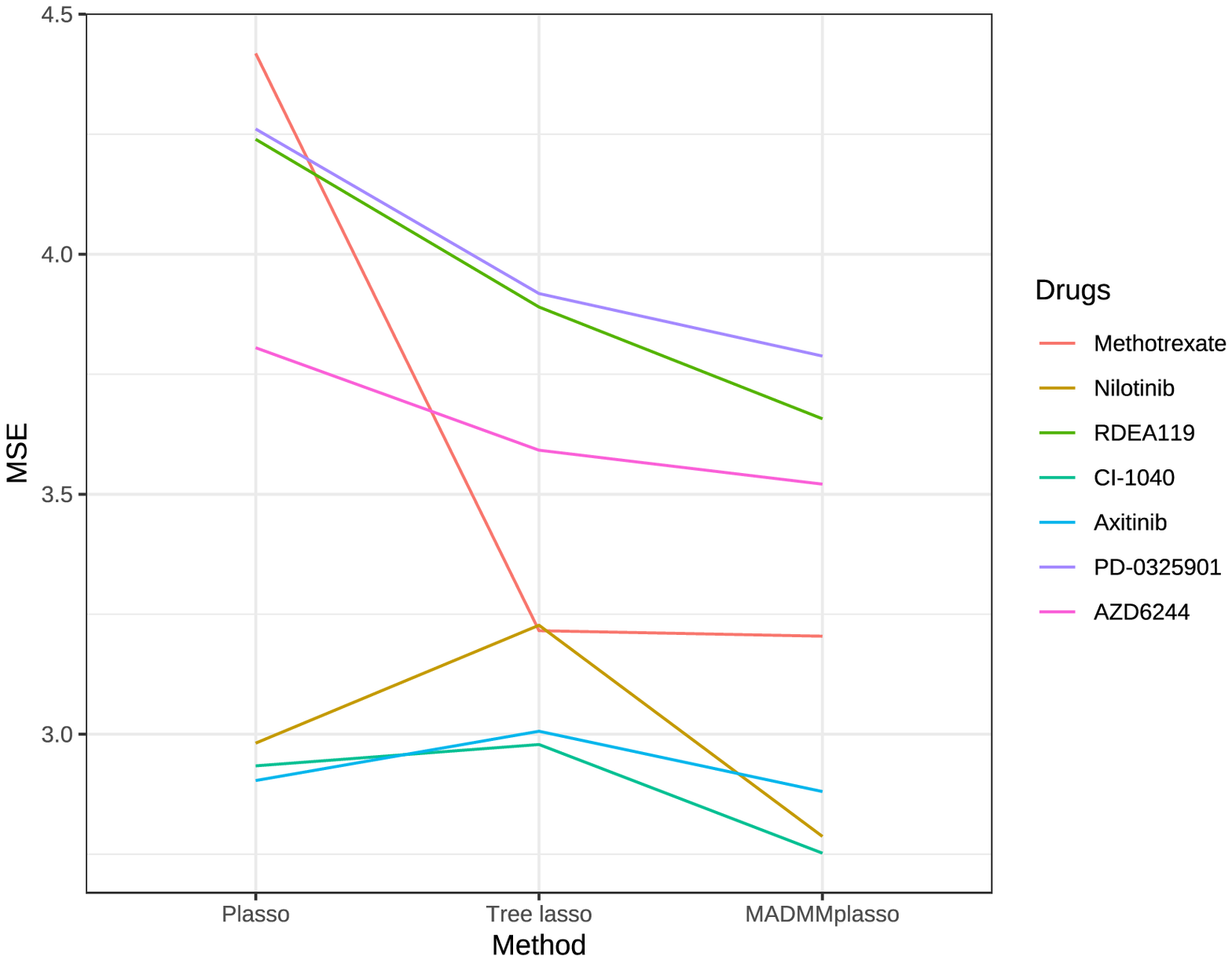}
 \caption{}\label{gdsc_mse}
\end{subfigure}
\caption{The general structure and the scatter plots  for the seven drugs, as estimated with complete-linkage hierarchical clustering using the Euclidean distance (a),  box plots of the overall mean squared error (MSE) on the test data (b) and the MSE for the individual drugs on the test data (c). The dendrogram (a) shows the hierarchical structure of the multivariate responses generated by the drugs $Y$. Below each node of the dendrogram is the name of the drug associated with the node.  The MSE on the test data ((b) and  (c)) were averaged across the 10 repeated data splits.}\label{tree_data}
\end{figure}

We obtained this data set from the archives of the GDSC database (ftp://ftp.sanger.ac.uk/pub4/cancerrxgene/releases/release-5.0/) \citep{10.1093/nar/gks1111}. The data set contains 97 cancer drugs tested on 498 cell lines. The cell lines are representations of 13 cancer tissues. The drug responses were summarized by the logarithm of the half maximal inhibitory concentration, i.e.  $\log(\text{IC}_{50})$, estimated from the dose-response curves that measured drug response at specific drug concentrations in terms of normalized relative cell viability. The data set also contains gene expression features and other genomic data such as DNA copy numbers and mutations.  In our experiment, we preselected 2602 gene expression features that explain  $50\%$ of the variation as $X$ and represented the cancer types in a 12-column matrix of dummy variables as $Z$. We further selected seven drugs for illustrative purpose. The structure of the drugs is shown in figure \eqref{tree3}. 

We randomly split the data into $80\%$ training and $20\%$ testing data and a high-dimensional multi-response regression model was fit using the training data, where $Y$ was the $\log (\text{IC}_{50})$ values. We performed this procedure repeatedly on 10 random splits and present the average of the results. We show the efficiency our model by comparing with the tree lasso (using the mixlasso R package)  and the pliable lasso. To allow the tree lasso to include the modifying variable, we allowed the cancer types $Z$ to be included without any penalization. To use the pliable lasso, we modeled each response $Y_d, \quad d=1,\ldots, D$ separately.  The summary of the results for the 10 random splits is shown in table \eqref{table3} and in figures  \eqref{Box4} and \eqref{gdsc_mse}. 
We find that the proposed model performs best in terms of prediction error on the test data. The proposed model is able to combine the information from the various cancer types  with corresponding gene expression features to determine which drug will be suitable for each cancer type. It also borrows statistical strength from the correlations that exist among the responses. This is very essential in the area of drug discoveries where it important to identify individual drugs or groups of drugs that could be helpful in treating specific cancers and subtypes of cancers.

\begin{table}[ht]
  \centering
    \caption{Results from the GDSC data.}
    \label{table3}
    \begin{threeparttable}
    \begin{tabular}{|l l l |} 
    \hline
     \textbf{Model} &  Non-zero coefficients\tnote{1} & Test error (SD)\tnote{2} \\
      \hline
     Plasso & 724 & 3.648 (0.270) \\
     Tree lasso & 1016 & 3.404 (0.268)  \\
MADMMplasso & 1424 & 3.227 (0.267)  \\      
     \hline
    \end{tabular}
    \begin{tablenotes}
    \item[1] The number of non-zero coefficients in the model. We counted  the coefficients with at least two non-zero values across the 10 repeated data splits. Number of non-zero coefficients $=\sum_{j=1}^p\sum_{d=1}^D\{(\sum_{r=1}^{10} \bm1_{\{\beta_{jd}^r\neq 0\}})\geq 2\}$

    Note that the selection is out of $p\times D=18844$  features in total. 
    \item[2] The MSE on an independent test data. We included the standard deviation (SD) across the 10 repeated data splits.
  \end{tablenotes}
    \end{threeparttable}
\end{table}

\begin{figure}[ht]

  \centering
  \includegraphics[width=\textwidth]{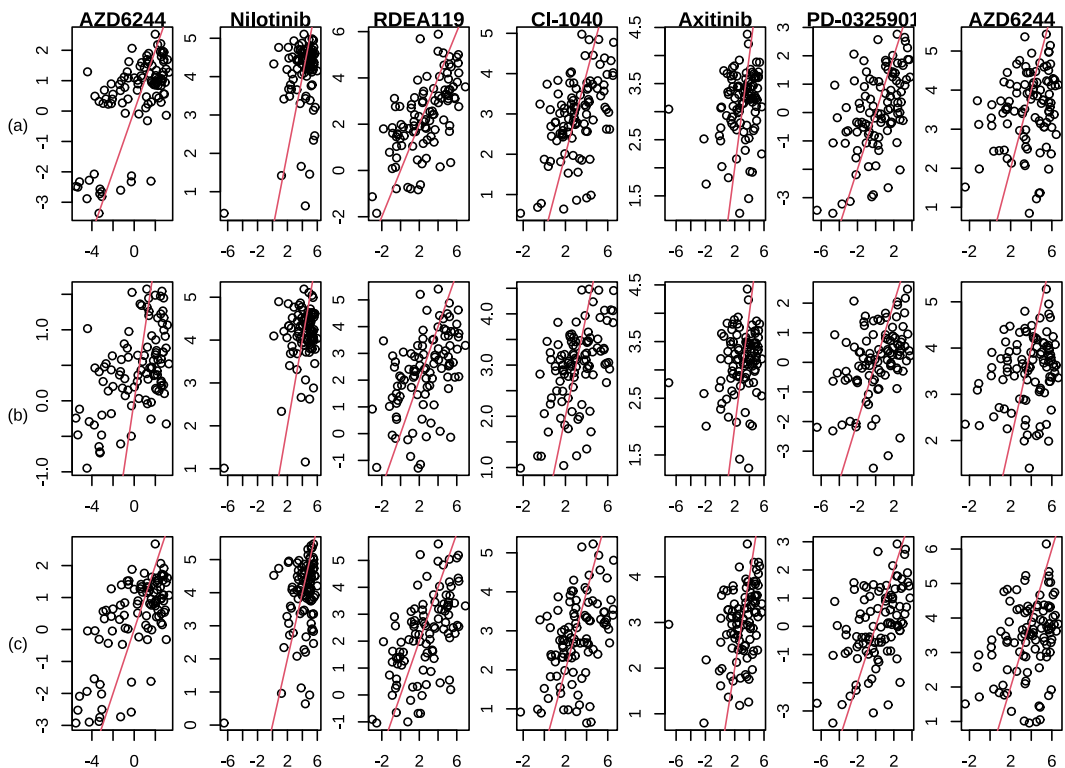}
\caption{Scatter plots on the observed drugs responses versus the predicted drug responses for MADMMplasso (a), plasso (b) and tree lasso (c). Note that these results were obtained for one of the random data splits. The x-axis  represents the observed responses and and y-axis  represents the predicted responses.}\label{scatter1}
\end{figure}

In figure \eqref{scatter1}, we show the prediction performance of the three models. We can see from the plots that all the models perform averagely well in terms of predicting the individual response variables $Y$. However, by looking at figure \eqref{gdsc_mse} which gives the averages of the test errors for the individual drugs we see that our proposed approach performs best in terms of prediction error. 

\begin{figure}[ht]
\centering
\begin{subfigure}{.4\textwidth}
  \includegraphics[width=7cm, height=6cm]{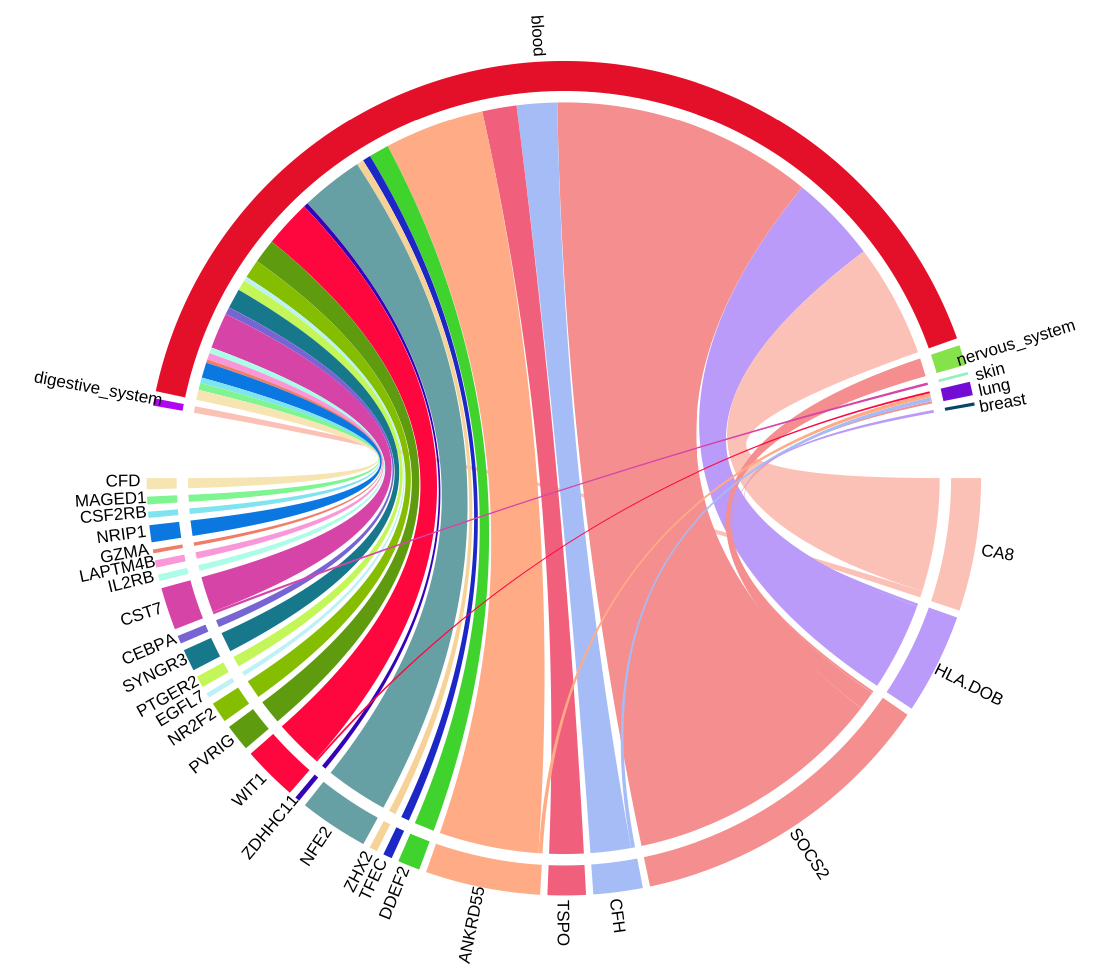}
 \caption{ }
\end{subfigure}
\hfill
\begin{subfigure}{.4\textwidth}
  \includegraphics[width=7cm, height=6cm]{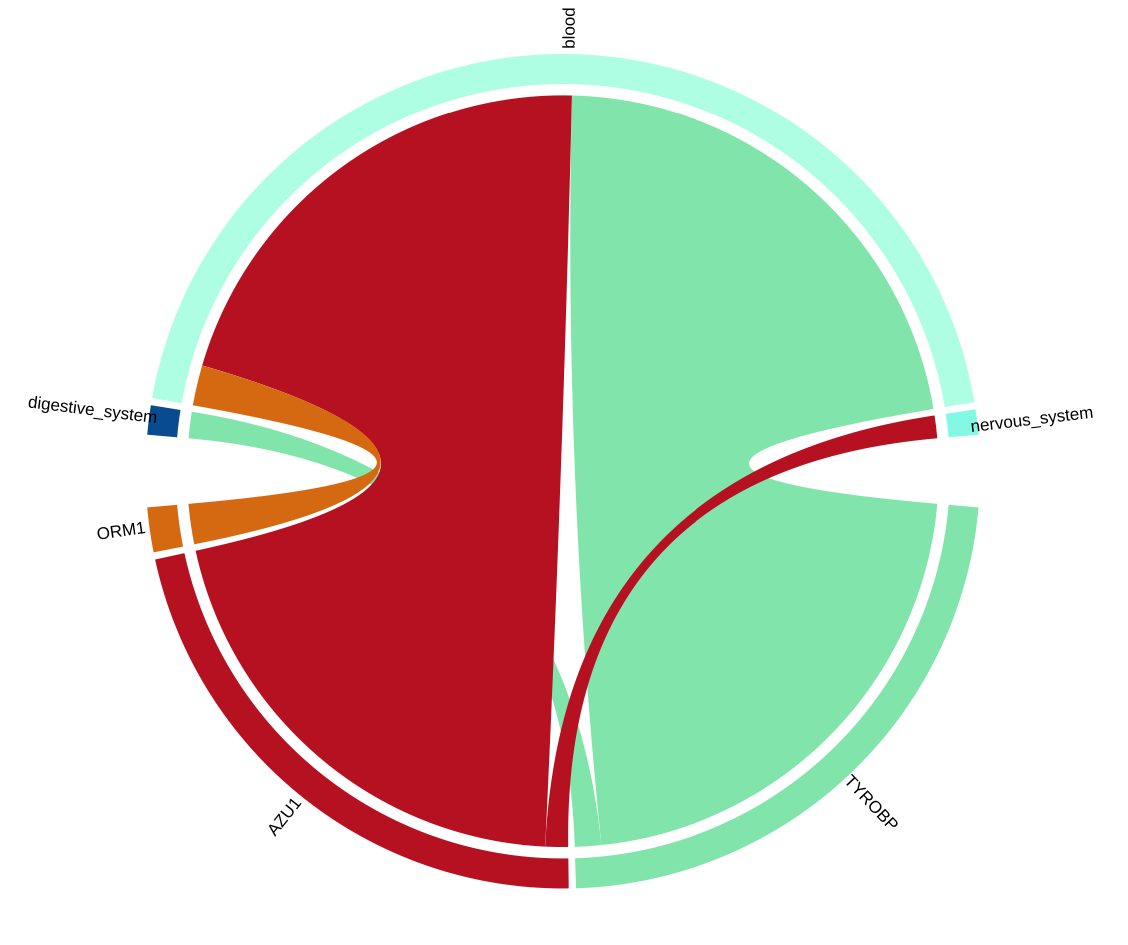}
 \caption{}
\end{subfigure}
\begin{subfigure}{.4\textwidth}
  \includegraphics[width=6cm, height=6cm]{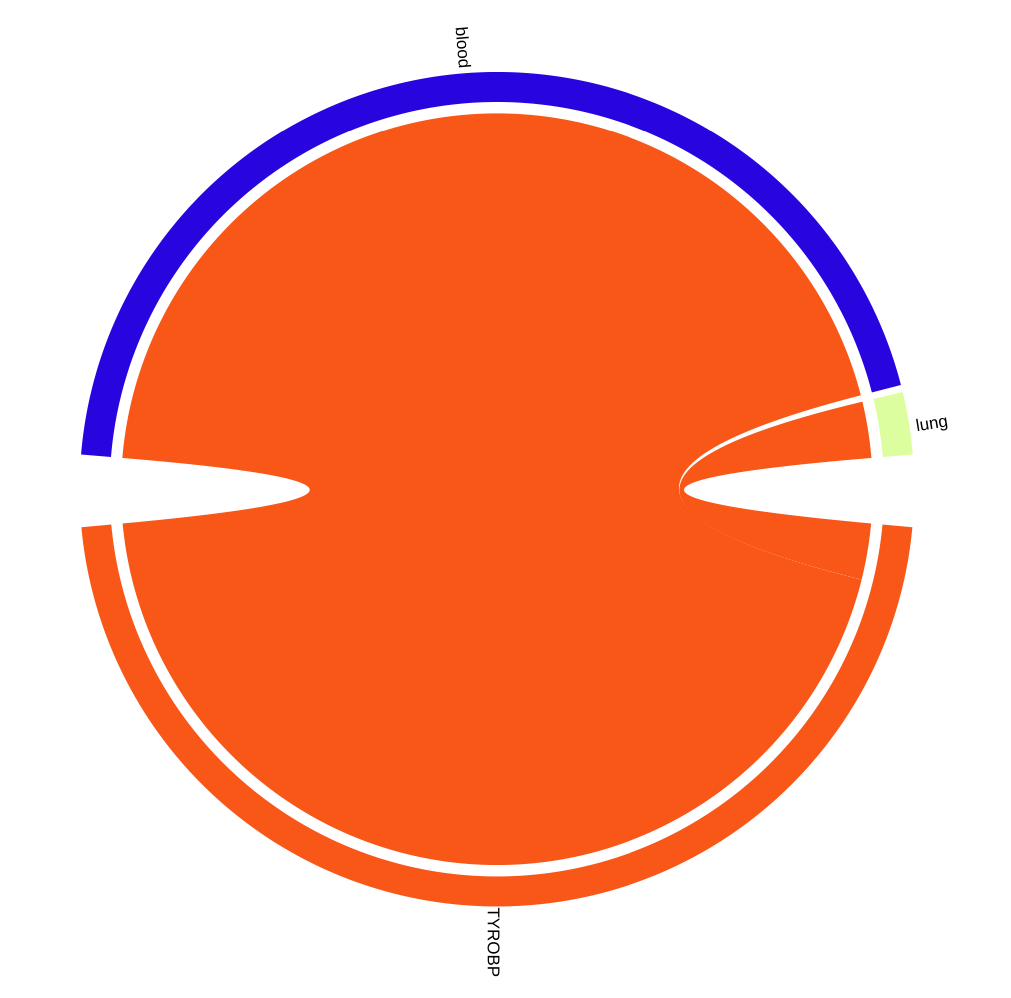}
 \caption{ }
\end{subfigure}
\hfill
\begin{subfigure}{.4\textwidth}
  \includegraphics[width=6cm, height=6cm]{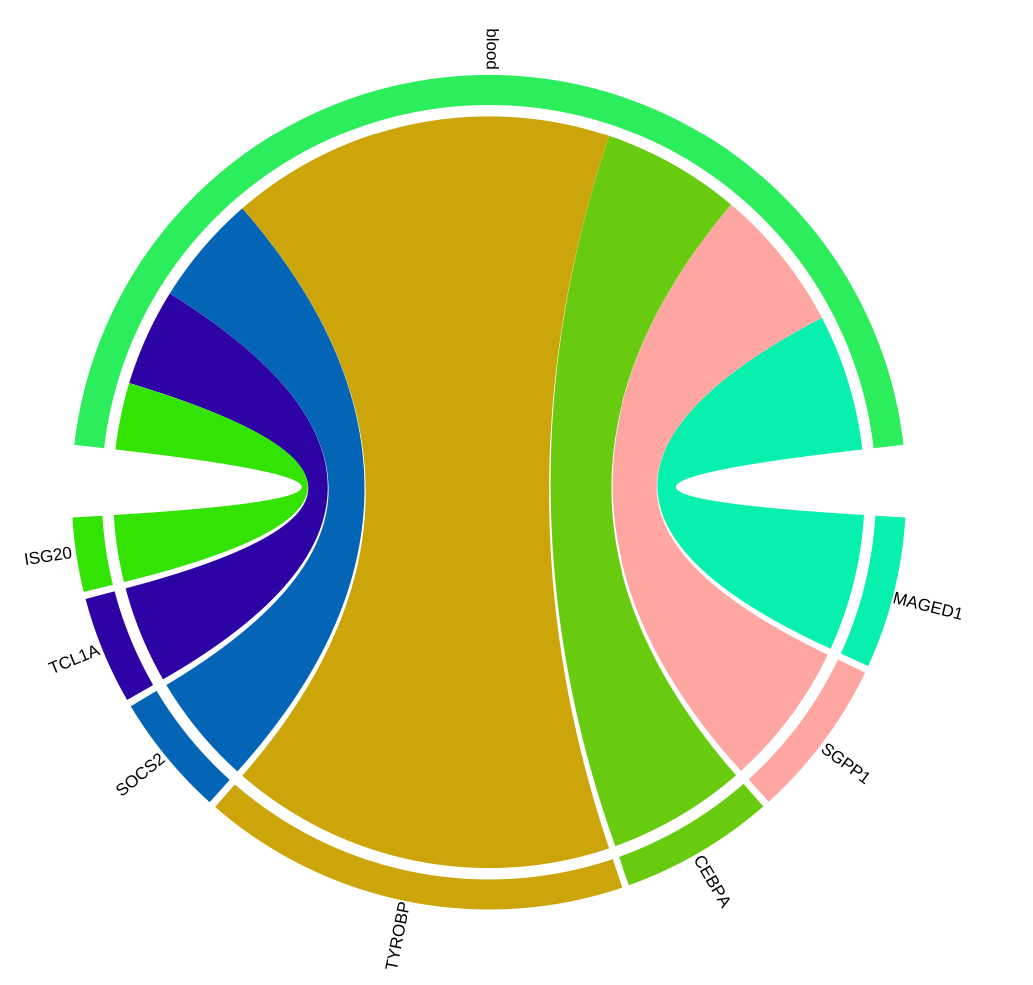}
 \caption{ }
\end{subfigure}
\begin{subfigure}{.4\textwidth}
  \includegraphics[width=7cm, height=6cm]{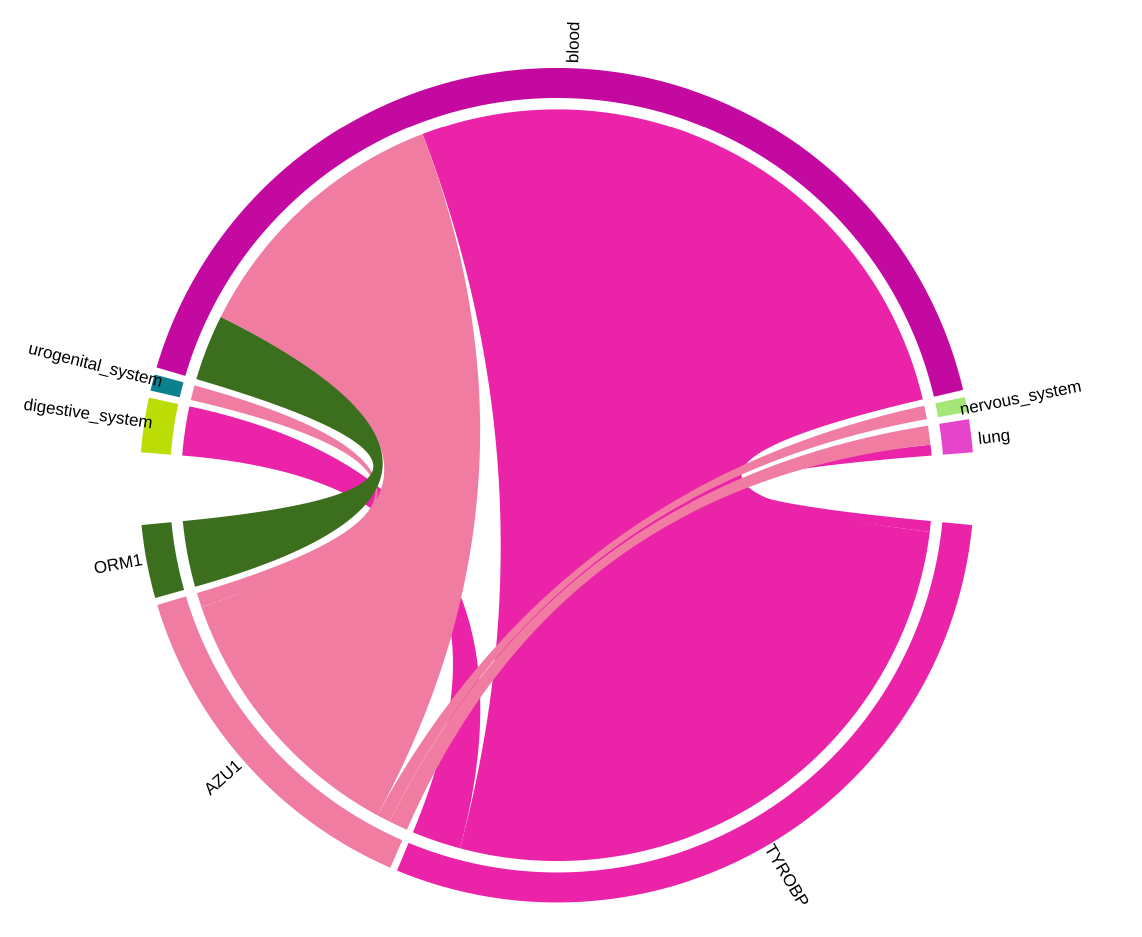}
 \caption{ }
\end{subfigure}
\caption{Chord diagrams for the interaction effects between gene expression features and cancer types for Nilotinib (a), RDEA119 (b), CI-1040 (c), Axitinib (d) and PD-0325901 (e). The upper chord represents the cancer types and the lower chord represents the gene expression features. The width of the line connecting the cancer type to a gene indicates the strength of the interaction. The selection was done for the coefficients with at least two non-zero values across the 10 random data splits. }\label{circplot_data}
\end{figure}

\begin{figure}[ht]
\centering
  \includegraphics[width=\textwidth]{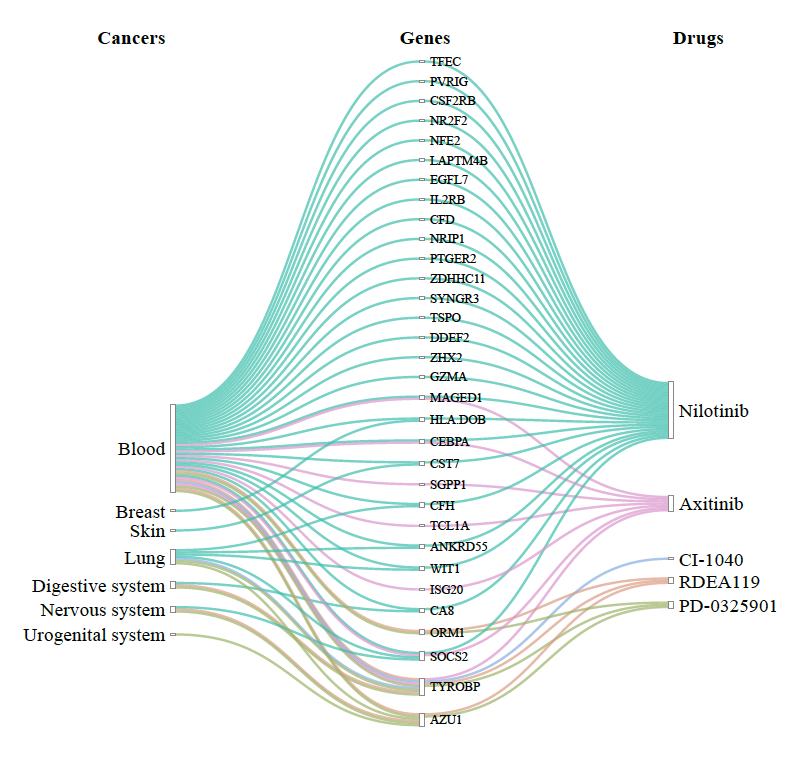}
\caption{The interaction plot (Sankey diagram) for the GDSC dataset. Each gene expression feature in the figure has at least one corresponding cancer type and at least one corresponding drug and each line connecting the three variables indicates a selected interaction effect between the gene expression feature and the cancer type for the given drug response variable. The lines have been been grouped according to the drugs. Note that the selection into the figure was done for the coefficients with at least two non-zero values across the 10 random data splits.   }\label{sankey_gdsc_data}
\end{figure}

Further analyses on the proposed model are shown in figures  \eqref{circplot_data} (for individual drugs) and \eqref{sankey_gdsc_data}\footnote{This was generated by using the R package ``EnrichIntersect" created by \citeauthor{enrichment},\citeyear{enrichment}.} (for all drugs together). These plots show the selected interactions between gene expression features and cancer types. We show the relationship between the identified genes (those considered to  interact with cancer types), the cancer types and the associated drugs, identified by the model to be linked to specific cancer types. We note here that the gene expression features shown in the plots are only those that interact with at least one  cancer type. It can be seen that not all the seven drugs were identified to be linked to specific cancer types. Specifically, methotrexate was not found by the model to be associated with a specific cancer type as no interactions between cancer type and gene expression features were selected in the MADMMplasso model for this drug. This is as expected as methotrexate is a generic chemotherapy agent that does not target specific molecular features or pathways in the cell. One of the interesting drugs identified is nilotinib which is associated with blood cancers, specifically it is a is a Bcr-Abl tyrosine kinase inhibitor which targets chronic myelogenous leukemias (CML) with the Philadelphia chromosome. A common gene associated with blood cancers as identified by all the three models is the suppressor of cytokine signaling 2 (SOCS2). In figure \eqref{circplot_data}, we can see that this gene is seen to interact with the blood cancer type. In fact,SOCS2 is highly expressed in many of the blood cancer cell lines in this data set. Studies have shown that SOCS2 is involved in the signal transduction cascades in CML cells \citep{10.1182/blood.V99.5.1766}. From the GDSC data set, we also realized that most of the cell lines with high expression of SOCS2 and having low score of $\log (\text{IC}_{50})$ for nilotinib are CML tumors. This provides evidence that the proposed model can help in identifying genes that are associated with diseases and their corresponding drugs in drug discoveries. 

We see from figure \eqref{sankey_gdsc_data} that the total number of identified interactions is quite small. Most interactions are observed for the blood cancer type. Blood cancers are biologically quite different from solid tumours, and the interaction effects allow the MADMMplasso model to modify the gene expression effect estimates to allow different effects for blood cancer compared to the solid tumours. Most interaction effects with the blood cancer type are identified for the nilotinib and axitinib drug response variables. This makes sense as both these drugs are Bcr-Abl tyrosine kinase inhibitor which are only applied to chronic myelogenous leukemias with Bcr-Abl mutations.

\clearpage
\section{Conclusion}\label{section6}
In this paper, we proposed a novel regularized regression method, which we named the MADMMplasso, to represent a multi-response pliable lasso regression model in combination with an ADMM algorithm for model fitting. This method is able to find covariates $X$ and their corresponding interactions with modifying variables $Z$, with some joint association with multiple related responses. The joint association takes into consideration the correlational structure in the response variables, assumed to form some overlapping groups, and specifically in this paper a hierarchical tree structure. We allowed the interaction term to be included in an asymmetrical weak hierarchical manner by first considering whether the corresponding main term is in the model. For optimization, we implemented an ADMM algorithm that allowed us to model the overlapping groups in a simple way. The results from the simulations and from the application to a pharmacogenomic screening dataset show that the ability of the proposed method to account for correlated responses and to include interaction effects can result in a clear improvement in prediction and variable selection performance. An implementation of the package is available on Github through the link https://github.com/ocbe-uio/MADMMplasso and the version used throughout this paper is 1.0.0.

\section*{Acknowledgements}
This work received funding from the European Union’s Horizon 2020 Research and Innovation program, under the Marie Skłodowska-Curie Actions Grant, agreement No. 801133 (Scientia fellowship), and under grant agreement No. 847912 (``RESCUER").

%
%

\bibliographystyle{abbrvnat} 
    \bibliography{references}
\end{document}